\documentclass[11pt,a4paper]{article}
\pdfoutput=1  
\usepackage{epsfig}
\usepackage{graphicx,psfrag}
\usepackage{multirow}
\usepackage{slashed}
\usepackage{pstricks}
\usepackage{caption}
\usepackage{subcaption}
\usepackage{url}
\usepackage{braket}
\usepackage{jheppub}
\usepackage{enumerate}
\usepackage{wasysym} 
\usepackage{mathrsfs} 
\usepackage{amsfonts} 
\usepackage{amsbsy} 
\usepackage{amscd}
\usepackage{color}

\makeatletter

\def\eeq{\end{equation}}
\def\beq{\begin{equation}}

\newcommand{\Rmnum}[1]{\expandafter\@slowromancap\romannumeral #1@}

\newcommand{\gsim}{\raisebox{-0.13cm}{~\shortstack{$>$ \\[-0.07cm]
      $\sim$}}~}
\newcommand{\lsim}{\raisebox{-0.13cm}{~\shortstack{$<$ \\[-0.07cm]
      $\sim$}}~}
\makeatother

\title{LHC signals of triplet scalars as dark matter portal: cut-based approach and improvement with gradient boosting and neural networks }

\author[a]{Atri Dey,}
\author[a]{Jayita Lahiri,} 
  \affiliation[a]{Regional Centre for Accelerator-based Particle Physics,
Harish-Chandra Research Institute, HBNI,
Chhatnag Road, Jhunsi, Allahabad - 211 019, India}

\author[b]{Biswarup Mukhopadhyaya} 
  \affiliation[b]{Indian Institute of Science Education and Research Kolkata, Mohanpur 741246, India}

\emailAdd{atridey@hri.res.in}
\emailAdd{jayitalahiri@hri.res.in}
\emailAdd{biswarup@iiserkol.ac.in}

\abstract{
We consider a scenario where an SU(2) triplet scalar acts as the portal
for a scalar dark matter particle. We identify regions of the parameter space,
where such a triplet coexists with the usual Higgs doublet consistently
with all theoretical as well as neutrino, accelerator and dark matter constraints, and the
triplet-dominated neutral state has substantial invisible branching fraction.
LHC signals are investigated for such regions, in the final state
{\em same-sign dilepton + $\ge$ 2 jets + $\not E_T$.} While straightforward
detectability at the high-luminosity run is predicted for some benchmark
points in a cut-based analysis, there are other benchmarks where 
one has to resort to gradient boosting/neural network techniques
in order to achieve appreciable signal significance.

}

\preprint{HRI-RECAPP-2020-001\\$\textrm{}$}

\begin{document}

\maketitle

\newpage

\section{Introduction}

The recent data on direct search for dark matter (DM), especially those from
the Xenon1T observation~\cite{Aprile:2018dbl}, rather strongly constrain scenarios where the
125 GeV Higgs acts as dark matter portal. The coupling of, say, a scalar SU(2) singlet
DM to the Higgs boson of the standard model (SM) is restricted by such constraints
to be $\lsim 10^{-3}$. Ensuring the DM annihilation rate required
for consistency with the observed relic density becomes a big challenge in
such a case. 

The restriction is considerably relaxed for an extended electroweak symmetry 
breaking sector. For example, in two-Higgs doublet models (2HDM), one can
have regions in the parameter space where the DM candidate has rather feeble
interaction with $h$, the SM-like scalar, but sufficient coupling with the
heavier neutral scalar $H$ so as to be consistent with both direct search results 
and the relic density~\cite{Drozd:2014yla}. This can happen due to the large mass of a mediating
$H$ suppressing the elastic scattering rates; it is also possible to have
cancellation between the $h$ and $H$-mediated scattering amplitudes. The allowed
regions in 2HDM satisfying such requirements and the corresponding signals
at the Large Hadron Collider (LHC) have been studied in detail~\cite{Dey:2019lyr}. 

Here we present the results of a similar investigation in the context of scalar triplet extension of the SM~\cite{Magg:1980ut,Schechter:1980gr,Lazarides:1980nt,Mohapatra:1980yp,Cheng:1980qt,Bilenky:1980cx,Kobzarev:1980nk,Gunion:1989ci,Mukhopadhyaya:1990up,Mohapatra:1991ng,Ma:1998dx,Chaudhuri:2013xoa,Chaudhuri:2016rwo}, together with a scalar DM candidate
$\chi$~\cite{Krolikowski:2012jd}. A $Y=2$ scalar triplet added to the SM doublet, with the triplet
vacuum expectation value (VEV) limited to $\lsim 4.8$ GeV by the $\rho$-parameter~\cite{Primulando:2019evb},
can generate Majorana masses for neutrinos via the Type-II Seesaw mechanism.
It also has a rich collider phenomenology~\cite{Godbole:1994np,Cheung:1994rp,Ghosh:1996jg,Perez:2008ha,Du:2018eaw,Ghosh:2018jpa}, largely due to the presence of
a doubly charged scalar that decays either to same-sign dileptons or
same-sign $W$ pairs. 

If the DM particle $\chi$, odd under a $Z_2$ symmetry, couples to such
a triplet $\Delta$, the strength of the interaction is not subject to
severe constraints. This is because the triplet cannot mediate the
elastic scattering of $\chi$ against the quarks in a terrestrial
detector, because of electroweak gauge invariance. On the other hand,
the SM-like scalar doublet must again have suppressed interaction with
$\chi$. The requisite DM annihilation rate in such a case can be
ensured by an appropriate quartic interaction $\Delta^\dagger \Delta\chi^2$,
on which no severe constraint exists. We have indeed found a
substantial region in the parameter space, satisfying all constraints
from direct search, relic density, neutrino masses and mixing, and of
course collider searches for triplet scalars. We thereafter look for
the LHC signals of such a scenario serving as DM portal, one of whose
consequences is an invisible branching ratio for $H$, the physical
state dominated by the neutral CP-even
member of $\Delta$.  This can be utilised in Drell-Yan (DY) processes
involving the doubly charged scalar. The most profitable DY channel
is found to be $pp \rightarrow H^{\pm \pm} H^{\mp}$, where $H^{\pm \pm}, H^{\mp}$
are once more the doubly- and singly-charged mass eigenstate dominanted
by components of the triplet. The $H^{\mp}$ in such a situation decays into $H W^{\mp}$;
we latch on to the invisible decay of the $H$ into a DM pair, while the $W$ is
identified in its hadronic decay channels.

The $\Delta L = 2$ lepton Yukawa interactions of $\Delta$ generate
neutrino masses. This puts constraints on the products of the triplet
VEV multiplied by the Yukawa couplings strengths. When the VEV is
small, relative large $\Delta L = 2$ interactions make same-sign
dileptons the dominant decay products of $H^{\pm \pm}$. In the other
situation, namely, one where the triplet VEV is close to its
experimental limit, this VEV drives the decay to $W^{\pm} W^{\pm}$ to be the
principal mode. We find that the first scenario has especially clean
signals, with large missing-$E_T$ ($\slashed{E_T}$) from invisible $H$-decay
accompanied by a sharp dilepton mass peak. The event rate in
vector boson fusion (VBF) channel is also estimated but found to be
inadequate for detection of the signal. Lastly, we follow up of the cut-based analysis with a
multivariate analysis based on gradient boosting, and also using the
artificial neural network (ANN) technique.

The plan of this work is as follows. In Section~\ref{sec2}, we present a brief outline of the model. In Section~\ref{sec3}, we discuss all the relevant constraints on our model including those from Higgs sector, dark matter sector, electroweak presicion tests, neutrino data and theretical constraints. In Section~\ref{sec4}, we choose appropriate final states and discuss interesting benchmark points for collider studies involving the model. In Section~\ref{sec5}, we present the cut-based collider analysis for selected final states. In Section~\ref{sec6}, we explore the scope for improvements using sophisticated neural network and gradient boosting analysis. We summarize our key findings of this work and conclude in Section~\ref{sec7}.

\section{A model with a triplet scalar and a scalar dark matter}
\label{sec2}
We concentrate on an extension of a Type-II Seesaw scenario containing a $Y$ = 2 scalar triplet $\Delta$ along with a singlet scalar dark matter candidate $\chi$. $\chi$ interacts with $\Delta$ and the SM-like higgs doublet $\Phi$ via terms in the scalar potential. The Lagrangian of the full scenario is

\begin{equation}
{\cal L} = {\cal L}_{SM} + {\cal L}_{Type-II Seesaw} + {\cal L}_{DM} + {\cal L}_{Int} 
\end{equation}
where
\begin{equation}
{\cal L}_{DM} + {\cal L}_{Int} = \frac{1}{2} \partial^{\mu} \chi  \partial_{\mu} \chi - \frac{1}{2} M_{\chi}^2 \chi^2 + \lambda_S \chi^4 + \lambda_{D} \chi^2 \Phi^{\dagger} \Phi + \lambda_{T} \chi^2 Tr(\Delta^{\dagger} \Delta)
\label{lagdm}
\end{equation}
\noindent
$\chi$, an $SU(2)_L \times U(1)_Y$ singlet, does not have any vacuum expectation value (VEV). An additional $Z_2$ symmetry ensures this, under which $\chi$ is assumed to be odd but $\Phi$ and $\Delta$ are even.
The $Z_2$ prevents $\chi$ from mixing with $\Phi$ and $\Delta$. Thus the phenomenological constraints on all particles/interactions except those involving $\chi$ are similar to those applicable on a Type-II Seesaw model.\\\\

{\bf The scalar potential of Type-II Seesaw model:}\\

The most general Higgs potential involving $\chi$, $\phi$ and $\Delta$ can be
written as 
\begin{eqnarray}
\label{lambdapotential}
\mathcal{V}(\Phi,\Delta,\chi) &=& a \Phi^\dagger\Phi+ \frac{b}{2} Tr(\Delta^\dagger\Delta) - \frac{1}{2} M_{\chi}^2 \chi^2 + c (\Phi^\dagger\Phi)^2 + \frac{d}{4} 
\left( Tr(\Delta^\dagger\Delta) \right)^2 \nonumber\\
&+& \frac{e-h}{2} \Phi^\dagger\Phi Tr(\Delta^\dagger\Delta)
+ \frac{f}{4} Tr(\Delta^\dagger\Delta^\dagger)Tr(\Delta\Delta) 
+ h \Phi^\dagger \Delta^\dagger \Delta \Phi  
+ ( t \Phi^\dagger \Delta \tilde{\Phi} + h.c )\nonumber\\
&+& \lambda_S \chi^4 + \lambda_{D} \chi^2 \Phi^{\dagger} \Phi + \lambda_{T} \chi^2 Tr(\Delta^{\dagger} \Delta)\,. 
\end{eqnarray}
where, $\tilde{\Phi} \equiv i\tau_2 \Phi^\ast$. This scalar sector is expressed in terms of additional scalar triplet with usual scalar doublet 
\begin{equation}\label{vev}
  \Phi(1,2,+1) =
 \left(\begin{array}{c} \phi^+ \\ \phi^0 \end{array}\right)
\quad \mbox{and} \quad
\Delta(1,3,+2)  =
\left(\begin{array}{cc} \delta^+ & \sqrt{2}\delta^{++} \\ \sqrt{2}\delta^0 & -\delta^+ \end{array}\right) \,.
\end{equation}
The numbers in parentheses denotes their representation under SM Gauge group $SU(3)_C \times SU(2)_L \times U(1)_Y$.

The VEVs of the doublet and the triplet are given by

\begin{equation}\label{vev}
 \langle \Phi_0 \rangle =
 \left(\begin{array}{c} 0 \\ v_D \end{array}\right)
\quad \mbox{and} \quad
\langle \Delta_0 \rangle  =
\left(\begin{array}{cc} 0 & 0 \\ 0 & v_T\end{array}\right) \,.
\end{equation}

We concentrate now on the part of (Equation~\ref{lambdapotential}) involving $\Phi$ and $\Delta$ alone. All the parameters we choose are real, excepting $t$ which can be complex in general. Thus 
we write $t = |t|e^{i\gamma^{'}}$ and
$v_T = \omega e^{i\gamma}$ with $\omega \equiv |v_T|$. The orders of
magnitude for the other parameters in the potential are indicated as
\begin{equation}\label{oomagn}
a,\:b \sim v^2 \:;\quad
c,\: d,\: e,\: f,\: h \sim 1 \:; \quad |t|\ll v \,.
\end{equation}
where $v=\sqrt{{v_D}^2+2{v_T}^2}$.
The minimum of the potential expressed in terms as of the VEVs, is given by ~\cite{Grimus:1999fz}
\begin{eqnarray}
V( \langle \phi \rangle_0 , \langle \Delta \rangle_0 ) & = & 
\frac{1}{2} a {v_D}^2 + \frac{1}{2} b \omega^2 + \frac{1}{4}c {v_D}^4 +
\frac{1}{4}d \omega^4 \nonumber \\
&+& \frac{1}{4} (e-h) {v_D}^2 \omega^2 + 
{v_D}^2 \omega|t|\cos(\gamma^{'}+\gamma) \,. \nonumber \\
&& \label{pot-vev}
\end{eqnarray}
The minimization condition in terms of $( v_D, \omega, \cos(\gamma^{'}+\gamma))$ yield
$\gamma^{'} + \gamma = \pi$ or
\begin{equation}\label{tvtphases}
v_T = -\omega e^{-i\gamma^{'}} \quad \mbox{and} \quad v_T t = - \omega |t| \,.
\end{equation}
and
\begin{eqnarray}
a + c{v_D}^2 + \frac{e-h}{2} \omega^2 - 2|t|\omega &=& 0 \,, 
\label{min1}\\
b + d \omega^2 + \frac{e-h}{2} {v_D}^2 - \frac{|t|}{\omega} {v_D}^2 &=& 0 \,.
\label{min2}
\end{eqnarray}

with fields shifted with respect to the VEV's, one can write 
\begin{equation}\label{vev}
 \Phi =
 \left(\begin{array}{c} \phi^+ \\ \frac{v_D+\phi_r+i \phi_i}{\sqrt{2}} \end{array}\right)
\quad \mbox{and} \quad
\Delta  =
\left(\begin{array}{cc} \delta^+ & \sqrt{2}\delta^{++} \\ v_T +\delta+i\eta & -\delta^+ \end{array}\right) \,.
\end{equation}
After Spontaneous Symmetry Breaking(SSB) three Goldstone bosons are eaten up by the $W$ and the $Z$ bosons. Thus after diagonalizing the mass matrices, one is left with a doubly charge scalar  $H^{\pm \pm} \equiv \delta^{\pm \pm}$, a singly-charged scalar $H^{\pm}$ and two neutral scalars $h$ and $H$, along with a neutral pseudoscalar $A$. The corresponding mass eigenvalues are

\begin{eqnarray}
\label{mh}
m^2_{h} & \simeq & 
2c{v^2_D} + \frac{(e - h - 2q)^2}{2c - q} \omega^2 \,, \\
\label{mH}
m^2_{H} & \simeq &
q{v^2_D} - \left[ \frac{(e - h - 2q)^2}{2c - q} - 2d \right] \omega^2 \,, \\
\label{mA}
m^2_{A} & = & q({v^2_D} + 4 \omega^2) \,, \\ 
\label{mpp}
m^2_{H^{\pm\pm}} & = & (h+q){v^2_D} + 2 f \omega^2 \,, \\
\label{mp}
m^2_{H^{\pm}} & = & (q+\frac{h}{2})({v^2_D} + 2 \omega^2) \,
\end{eqnarray}

The diagonalization process also yields
\begin{eqnarray}
\label{tanalpha}
\tan \alpha & = & \frac{\sqrt{(q-2c)^2 {v_D}^2 + (2dq-4cd+(h-e+2q)^2)w^2}}{(h-e+2q)\omega} \,, \\
\tan \beta & = & \frac{2 \omega}{v_D}\,, \\
\tan \beta^{'} & = & \frac{\sqrt{2} \omega}{v_D}\,
\end{eqnarray}
where $\alpha$ is the mixing angle between the  CP-even parts of $\Phi$ and $\Delta$, $\beta$ is the mixing angle in charge Higgs sector with the mixing angle $\beta^{'}$ in the CP-odd Higgs sector. We can notice that only the CP-even scalars $h$ and $H$ can act as portal for dark matter where CP is conserved.\\

{\bf Gauge interactions:}\\ \\
The Gauge interaction terms are as usual as SM with additional term added for the triplet part

\begin{equation}
{\cal L}_{gauge} = (D_{\mu} \Phi)^{\dagger} D^{\mu} \Phi + \frac{1}{2} Tr((D_{\mu} \Delta)^{\dagger} (D^{\mu} \Delta))
\end{equation}
Where $D_{\mu} \Phi = \partial_{\mu} \Phi -   \frac{i}{2} g W_{\mu}^a \tau^a \Phi -  \frac{i}{2} g' B_{\mu} \Phi$ and $D_{\mu} \Delta = \partial_{\mu} \Delta -   \frac{i}{2} g [W_{\mu}^a \tau^a, \Delta] -  i g' B_{\mu} \Delta$ and $\tau^a$ are the SU(2) generators. \\

The gauge interactions will turn out be useful in our scenario where $\lambda_D \ll \lambda_T$ and thus the triplet scalar serves effectively as dark matter portal. As we shall see, we need to utilize the Drell-Yan production of triplet dominated states, driven by gauge couplings, for signals identifying the DM particle $\chi$.\\

{\bf Yukawa interactions:}\\ 

The triplet within this model have potential to induce Majorana neutrino masses via interactions with the left-handed lepton doublet $L \equiv (\nu,l)^T $ ~\cite{Perez:2008zc,Primulando:2019evb}. The Yukawa terms with $(\Delta L = 2)$ can be written as
\begin{equation}
{{\cal L}^{new}_Y} = \sqrt{2} f_{ab} {L_a}^T Ci\sigma^2 \Delta L_b + h.c.
\end{equation}
Where $C$ is the charge conjugation matrix and $a,b$ run over all three flavour indices. The neutrino masses are mostly dependent on the triplet VEV $w$ and can be expressed as
\begin{equation}
M_\nu = 2 f \omega
\end{equation}
As $f_{ab}$ is symmetric under $a\leftrightarrow b$, $M_{\nu}$ turns out to be a symmetric matrix. We can get the masses of the neutrinos after the diagonalization of $M_\nu$ with the help of the Pontecorvo-Maki-Nakagawa-Sakata (PMNS) matrix.\\

\section{Constraints and allowed regions of the parameter space}\label{sec3}

So long as there is small mixing between the dark matter particle
$\chi$ and the scalar triplet and doublet, which is ensured by the smallness
of the triplet VEV as compared to that of the doublet, the main constraints on the
scalar sector remain similar as for the Type-II Seesaw model, as
discussed in ~\cite{Primulando:2019evb}. We summarize them below, and turn to the additional
constraints on the dark matter sector.

It is useful to constrain the model parameters in terms of physical masses and mixing angles. Thus
we express the parameters in the potential as 
\begin{eqnarray}
\label{d}
d & = & \frac{1}{2 \omega^2}\left[ \frac{{m^2_h} \tan^2 \alpha + {m^2_H}}{1+\tan^2 \alpha} -q{v^2_D} \right] \,, \\
\label{c}
c & = & \frac{1}{2 v^2_D}\left[ \frac{{m^2_H} \tan^2 \alpha + {m^2_h}}{1+\tan^2 \alpha} \right] \,, \\
\label{q}
q & = & \frac{m^2_A}{v^2_D + 4 \omega^2} \,, \\
\label{h}
h & = & 2\left[ \frac{m^2_{H^{\pm}}}{v^2_D + 2 \omega^2} - \frac{m^2_A}{v^2_D + 4 \omega^2} \right]  \,, \\
\label{f}
f & = & \frac{1}{2 \omega^2}\left[ m^2_{H^{\pm \pm}} -  \frac{m^2_{H^{\pm}} v^2_D}{v^2_D + 2 \omega^2} \right] \,, \\
\label{e}
e & = & \sqrt{\frac{\left[ (q-2c)^2 v^2_D + 2(q-2c)d \omega^2 \right]}{(\tan^2 \alpha -1)\omega^2}} + 2q + h \,
\end{eqnarray}

Our adopted model has been encapsulated in a file  in {\bf Feynrules}
~\cite{Alloul:2013bka}. In our convention, the mixing angle $\alpha$
(Equation~\ref{tanalpha}) is such that $\alpha \rightarrow \pi/2$
aligns the lightest neutral scalar $h$ as the SM-like 125 GeV
Higgs. Equations~\ref{tanalpha}, \ref{mH} and \ref{mA} tell us that,
in the limit of small triplet VEV, $m_A$ and $m_H$ become nearly degenerate, which
is helpful in satisfying various constraints. 

\subsection{Constraints on relevant parameters of ${\cal L}_{Type-II Seesaw}$}\label{subsec3.1}
Theoretical constraints come mainly from the requirement of vacuum stability
and perturbativity at the TeV scale. We are not concerned
with ultraviolet completion here. In the expression for
the scalar potential in
Equation~\ref{lambdapotential}, all quartic terms involving just $\Phi$ and
$\Delta$ must be such that the scalar potential remains bounded
from below in any direction of the field space. 
The consequent vacuum stability conditions are ~\cite{Dey:2008jm,Akeroyd:2010je,Arhrib:2011uy,Bonilla:2015eha}

\begin{eqnarray}
\label{vs1}
4c \geq 0 \,, \\
\label{vs2}
\frac{d}{4} - f \geq 0 \,\\
\label{vs3}
\frac{e-h}{2} + \sqrt{4c(\frac{d}{4}-2f)} \geq 0 \,\\
\label{vs4}
\frac{e-3h}{2} + \sqrt{4c(\frac{d}{4}-2f)} \geq 0 \,\\
\label{vs5}
2f\sqrt{4c}+ |2h|\sqrt{(\frac{d}{4}-2f)} \geq 0 \,
\end{eqnarray} 

For perturbativity at the electroweak scale ~\cite{Cornwall:1974km,Dicus:1992vj}, one  demands that the 
quartic couplings at the EWSB scale must obey  
\begin{equation}
C_{H_i H_j H_k H_l} < 4\pi
\end{equation}
Where $C_{H_i H_j H_k H_l}$ include all quartic couplings.
Tree-level unitarity in the scattering of Higgs bosons and the
longitudinal components of the EW gauge bosons demands that the eigenvalues
of the scattering matrices have to be less than $16 \pi$ ~\cite{Arhrib:2011uy}.

Next come the phenomenological constraints. The two VEVs $v_D$ and
$w \equiv |v_T|$ decide the masses of $W\pm$ and $Z$, via
the expressions $m^2_W = g^2(v^2_D+2v^2_T)/4$ and $m^2_Z= (g^2+g^{'2})(v^2_D +
4v^2_T)/4$. Thus the ratio of these two gauge boson masses which is
constrained by the $\rho $ parameter, can be defined as $\rho \equiv m^2_W/(m^2_Z
\cos^2\theta_W) \equiv 1-\frac{2 v^2_T}{v^2_D + 4v^2_T}$. This puts an upper
bound on $|v_T|$, namely, $|v_T| \lsim 4.8$ GeV at 95\% CL.

Other constraints arise from
electroweak precision measurements, especially those of the oblique
parameters $S$ and $T$ ~\cite{Lavoura:1993nq,Chun:2012jw}. However, the augmentation
of the SM spectrum in terms of a scalar
triplet in general does not affect them seriously, as long as
the custodial SU(2) breaking is small. Loop contributions
to gauge boson self-energies remain within control with relatively
less effort, being suppressed by the square of the triplet VEV.
We refer the reader to reference~\cite{Chun:2012jw} for the derived $2\sigma$ limits
on the mass splitting between the triplet-dominated scalar mass
eigenstates, which has been obeyed in the regions of parameter
space used by us for the demonstration of our numerical results.

The LHC constraint on the heavy neutral scalar in such a scenario
consists of upper limits on the values of
$\sigma \times \text{Br}$ which can be translated to put some bound on
the parameter space ~\cite{Kanemura:2013vxa,Kanemura:2014goa}.
However, the experimental bound on $m^2_{H^{\pm \pm}}$ can be
easily determined from 95\% CL of $\sigma(pp \rightarrow
H^{++}H^{--})\times Br(H^{\pm \pm} \rightarrow \ell^{\pm} \ell^{\pm})$ ~\cite{Aaboud:2017qph},
in cases where the same-sign dilepton decay is the dominant channel
for the doubly charged scalar. The limit is much weaker~\cite{Aaboud:2018qcu} for 
high triplet VEV, when the $H^{\pm\pm}$ decays mostly into 
a same-sign $W$ pair. 
The choice of our benchmark points, as
discussed in the next section, takes these limits into account.

\subsection{Constraints on the dark matter sector}\label{subsec3.2}

As the scenario under consideration treats $\chi$ as a weakly interacting thermal dark matter candidate, 
it should satisfy the following constraints:
\begin{itemize}
\item The thermal relic density of $\chi$ should be consistent with the latest Planck limits
at the 95\% confidence level~\cite{Ade:2013zuv}. 

\item The $\chi$-nucleon cross-section should be below 
the upper bound given by XENON1T experiment~\cite{Aprile:2018dbl} and any other
data as and when they come up.

\item Indirect detection constraints coming from both isotropic gamma-ray data
and the gamma ray observations from dwarf spheroidal galaxies~\cite{Ackermann:2015zua} 
should be satisfied at the 95\% confidence level. This is turn
puts an upper limit on the velocity-averaged $\chi$-annihilation cross-section~\cite{Ahnen:2016qkx}.

\item The invisible decay of the 125-GeV scalar Higgs $h$ has to be $\leq$
15\%~\cite{Sirunyan:2018owy}. This includes contributions to both a $\chi$-pair
and any $\Delta L =2$ decay into neutrino pairs via doublet-triplet mixing.
\end{itemize}

The vacuum stability limits should not differ from those listed in
the previous susbsection, since $\chi$ represents a flat direction,
so far as the vacuum structure is concerned.  
In addition, perturbativity of all scalar quartic couplings demands
$0 < \lambda_S < 4 \pi$, $|\lambda_{D}| , |\lambda_{T}| < 4\pi$.

\subsection{The relevant parameter space}

We perform a wide scan of the model parameter space to identify
 regions which satisfy all the aforementioned constraints. Keeping in mind
scalar masses that are accessible to LHC searches, an exhaustive scan is 
contained in the following range choice:
\begin{center}
\label{scan_parameters}
$m_{\chi} \in \left[60,500\right]$ GeV, $m_H^{\pm} \in\left[100,1000\right]$ GeV, $m_H^{\pm \pm} \in \left[100,1000\right]$ GeV, \\ 

$|v_T|\equiv \omega \in \left[10^{-5},4.8\right]$ GeV, $|\sin \alpha | \in \left[0.999, 1\right]$ ,\\ 

$\lambda_{D} \in [-12,12]$, $\lambda_{T} \in [-12,12]$
\end{center}
Another important thing to notice is that the perturbativity conditions for $d$ and $f$ are quite 
sensitive to the mass eigenvalues of the triplet-dominated states, including their
splitting. With this as well as all precision constraints in view, our preferred benchmarks
are tilted towards regions corresponding to

\begin{eqnarray}
\label{relatedmasses}
m_A & \approx & \sqrt{\left( \frac{2 m^2_{H^{\pm}}}{v^2_D + 2 \omega^2} - \frac{m^2_{H^{\pm \pm}}}{v^2_D}\right)(v^2_D + 4 \omega^2)} \,, \\
m_H & \approx & m_A \,
\end{eqnarray}
with $\Delta m  =  m_H^{\pm} - m_H^{\pm \pm}$.

\begin{figure}[!hptb]
\centering
\includegraphics[width=12.0cm, height=10cm]{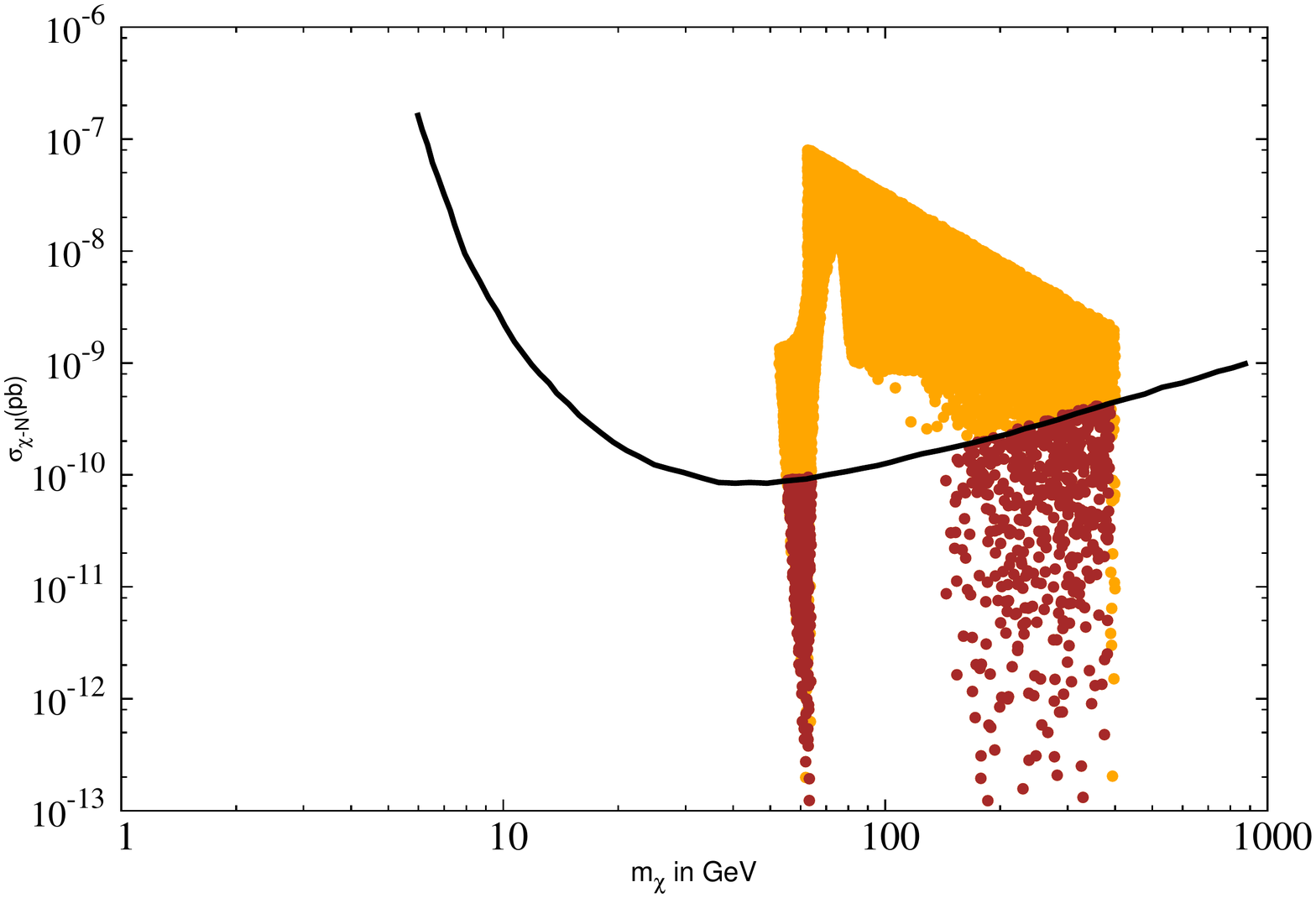} 
\caption{Parameter space allowed by the relic density observation. The black line is the upper limit on the $\chi-N$ scattering cross-section from XENON1T experiment. }
\label{omega_dd}
\end{figure}

Figure~\ref{omega_dd} represents a scatter plot generated from the scan, compared 
with the allowed region in the $m_{\chi} - \sigma_{(\chi - N)}$ space obtained from the current XENON1T data~\cite{Aprile:2018dbl}. 
The yellow region satisfies all constraints including those from relic density, while the black curve shows the upper limit on cross-section for spin-independent nucleon-DM scattering coming from XENON1T.
Note that the allowed region in the narrow strip in this figure corresponds to $m_\chi \simeq m_h/2$ and triplet VEV $\lsim 10^{-4}$ GeV. This is because
all other regions below the curve with such small triplet VEV, although allowed by direct searches, do not ensure the required
annihilation rate, unless one is close to the SM-like Higgs resonance. On the other hand, when the triplet VEV $\omega$ increases, the heavy CP-even state ($H$) starts contributing to the annihilation process. Therefore, regions with higher $m_{\chi}$ become allowed by the relic density requirements.

We use the global fit of neutrino data performed by the NuFITGroup ~\cite{Esteban:2018azc} (which
basically constrains the triplet VEV times the $\Delta L =2$ Yukawa interactions)  in zeroing
in on the benchmarks. We illustrate our results corresponding the case where all neutrino masses are nearly degenerate with the lightest neutrino mass $m_1 \approx 0.1$ eV. However,
the LHC-related prediction does not change appreciably (beyond 10\%) in the normal hierarchy (NH) or
inverted hierarchy (IH) scenarios as well. In the degenerate case, using the central values
of entries in the PMNS matrix~\cite{Primulando:2019evb}, one obtains 

\begin{equation}
M_\nu = 
\begin{pmatrix}
98.6 e^{i 0.0244} & 14.4 e^{-i 1.64} & 12.3 e^{-i 1.65} \\
14.4 e^{-i 1.64} & 106 e^{-i 0.0120} & 4.93 e^{-i 0.22} \\
12.3 e^{-i 1.65} & 4.93 e^{-i 0.22} & 104 e^{-i 0.0085} 
\end{pmatrix}
\end{equation}

As already mentioned, $M_\nu$ is fixed by neutrino oscillation data. 
We remind the reader that the same-sign dilepton channel for the doubly charged Higgs (which is 
a game-changer in collider signatures) is enhanced for small triplet VEV. For small $f_{ab}$,
on the other hand, the $W^{\pm} W^{\pm}$ decay channel dominates.

\section{Signals and benchmarks}\label{sec4}

Having identified the parameter space allowed by all constraints from
the Higgs sector and dark matter sector, we now proceed to look for
experimental probes for the scenario where the heavy neutral scalar $H$ of
Type II Seesaw model serves as DM portal. As the foregoing discussion
amply indicates, it is imperative to look at the invisible decay of $H$. The
production cross-sections of $H$ by both gluon fusion and vector boson
fusion(VBF) are suppressed by the factor $\frac{v_T^2}{v_D^2}$. The
Drell-Yan(DY) production of $H^{\pm\pm}H^{\mp}$ on the other hand
is driven purely by gauge couplings. We also mention here that the cross
section $\sigma(p p \rightarrow H^{\pm \pm} H^{\mp})$ increases with
large negative values of $\Delta m \equiv m_{H^\pm} - m_{H^{\pm    \pm}}$.
Keeping this in mind, we consider DY production of
$H^{\pm\pm}H^{\mp}$, followed by the $H^{\pm}$ decaying into $HW^{\pm}$ channel. 
The $H$, as we have seen, can decay invisibly with a substantial  branching
ratio, and thus gives rise to $\slashed{E_T}$. The $H^{\pm\pm}$ can decay into a
same-sign dilepton pair($\ell^{\pm}\ell^{\pm}$)~\cite{Aaboud:2017qph} or a pair or same-sign $W$
bosons ($W^{\pm}W^{\pm}$)~\cite{Aaboud:2018qcu}, depending on the value
of the $\Delta L = 2$ Yukawa interactions and the triplet VEV. These
two decay channels thus turn out to be complimentary
to each other, as will be discussed shortly.
    
\begin{figure}[!hptb]

\includegraphics[width=7.4cm, height=6.3cm]{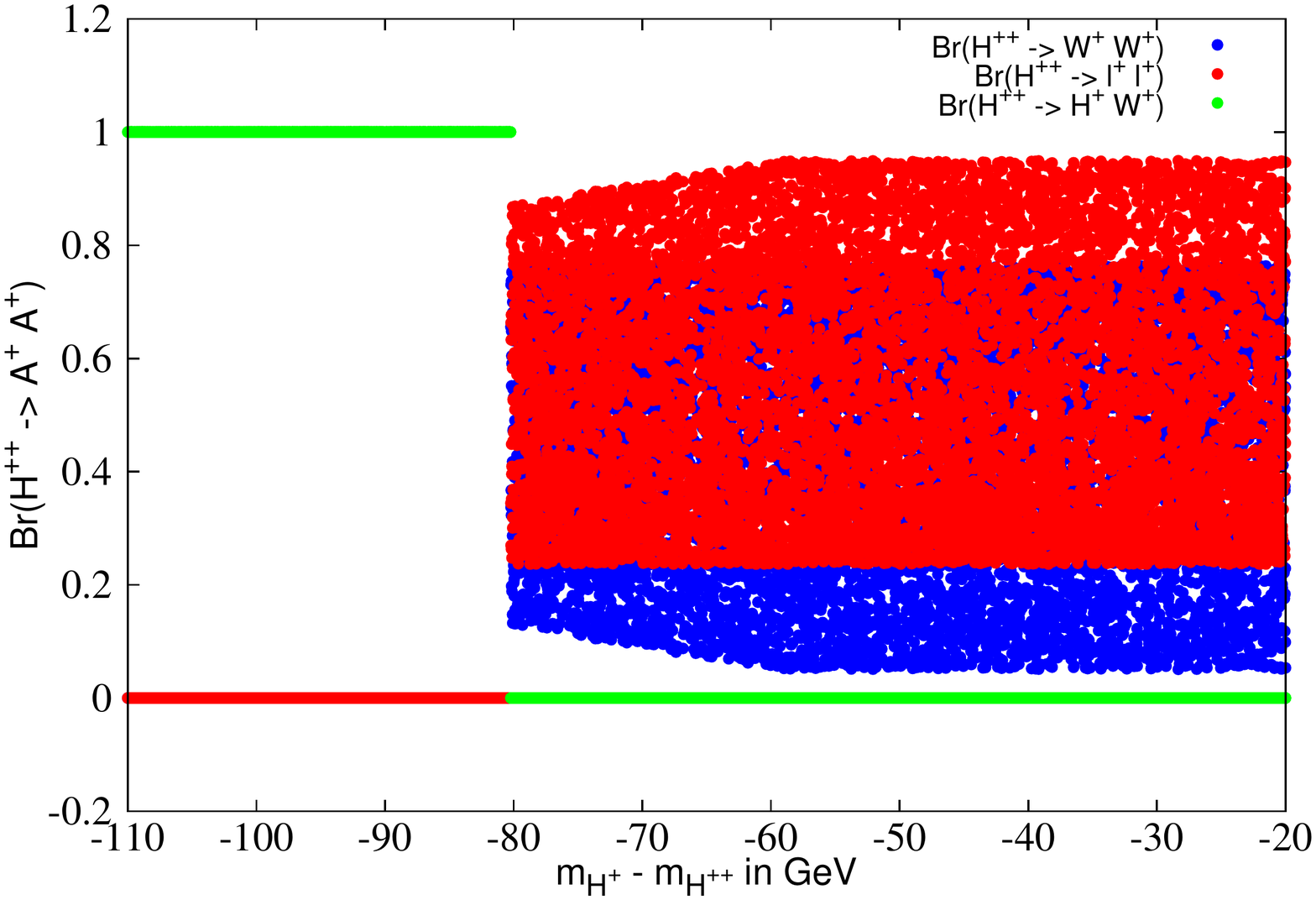}
\hspace{0.00001cm}
\includegraphics[width=7.4cm, height=6.3cm]{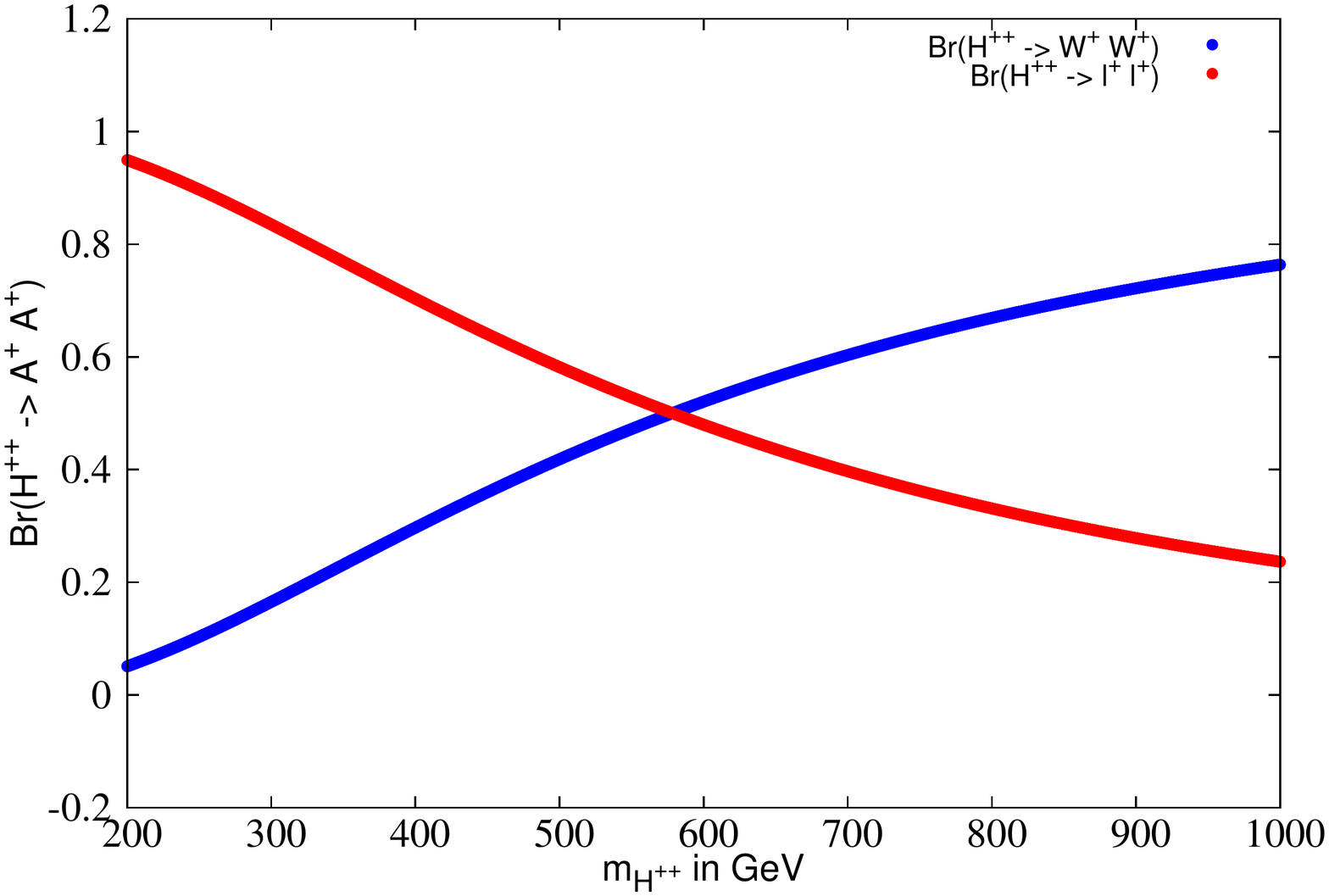}
\centering
\caption{Distribution of various branching ratios of $H^{\pm \pm}$ decays.}
\label{hppbrs}
\end{figure}

The choice of benchmark points in the parameter space, which will
highlight the efficacy of our signals, requires a little attention
to the important decay modes of
$H^{\pm\pm}$. In Figure~\ref{hppbrs} (left panel) we can see that as
long as $|\Delta m| \equiv |m_{H^\pm} - m_{H^{\pm \pm}}|$ is within 80
GeV, we can get sufficiently high branching fractions for $H^{\pm \pm}$
decay to $\ell^{\pm} \ell^{\pm}$ and $W^\pm W^\pm$ channels. As
soon as $|\Delta m|$ crosses 80 GeV, the channel
$H^{\pm \pm} \rightarrow H^{\pm} W^{\pm}$ opens up and dominates the decay.
However, SU(2) invariance of the theory, together with
the constraints from precision electroweak measurements
does not usually favour
such large mass splitting, when the triplet VEV is small, and
one has not more than one triplet. Thus we concentrate on
the scenarios corresponding to $H^{\pm\pm} \rightarrow \ell^\pm \ell^\pm$
and $H^{\pm\pm} \rightarrow W^\pm W^\pm$.
A very close degeneracy of the two charged physical states,
on the other hand, amounts to a suppression of the on-shell
$HW^{\pm}$ mode of the singly charged scalar.
The maximum mass splitting one finds compatible
with the above above constraints is $|\Delta m| \in [70, 80]$ GeV.

\begin{figure}[!hptb]
\centering
\includegraphics[width=12.0cm, height=8cm]{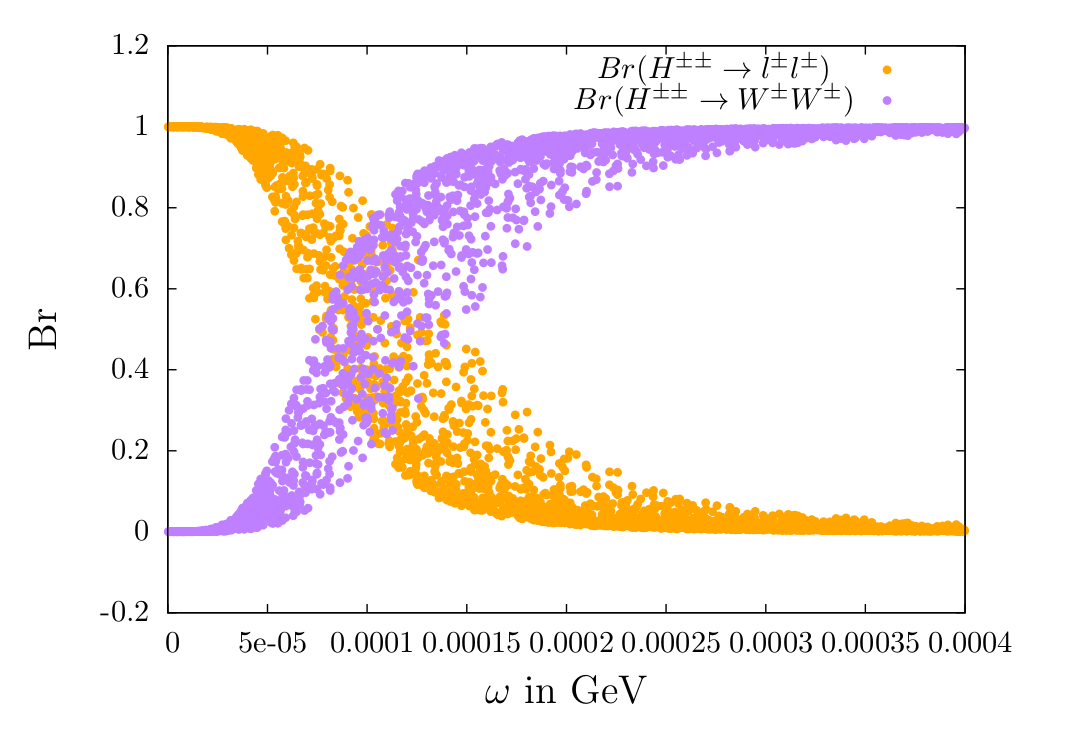}
\caption{Branching fractions of doubly-charged Higgs decaying into $\ell^{\pm}\ell^{\pm}$ and $W^{\pm}W^{\pm}$ channels as a function of triplet VEV. }
\label{llwwbr}
\end{figure}

Figure~\ref{llwwbr} shows the relative strengths of the two
channels as functions of the triplet VEV, the bands arising due
to the allowed ranges of the neutrino mass eigenvalues in
the NH scenario. One can see that, when the VEV of the triplet is 
$\le 10^{-5}$ GeV, $H^{\pm\pm}$ dominantly decays to $\ell^\pm \ell^\pm$. For
$w \gsim 10^{-4}$ GeV, on the other hand, the
$W^{\pm} W^{\pm}$ decay mode of $H^{\pm\pm}$ becomes dominant, as is evident
from Figure~\ref{llwwbr}. The phenomenology is strongly dependent on the fact that
the mixing angle($\alpha$) between the two CP-even neutral scalar states is
rather small, implying that $\sin \alpha \simeq 1$.

We have already seen that for $\omega < 10^{-5}$ GeV, $Br(H^{\pm\pm}
\rightarrow \ell^{\pm}\ell^{\pm}) \simeq 100\%$
while for $\omega > 10^{-4}$ GeV, $Br(H^{\pm\pm} \rightarrow W^{\pm}W^{\pm}) \simeq 100\%$.
In the intermediate
region they are comparable with each other, and the
branching ratio in either channel will depend on the mass of
the doubly-charged Higgs. The right panel in Figure~\ref{hppbrs}
describes the competition between the $Br(H^{\pm \pm} \rightarrow
\ell^{\pm} \ell^{\pm})$ and $Br(H^{\pm \pm} \rightarrow W^{\pm} W^{\pm})$ as
a function of $m_H^{\pm \pm}$ in such intermediate regions ($\omega
\sim 10^{-4}$ GeV). It can be clearly seen that as $m_{H^{\pm \pm}}$
increases it favours $W^{\pm} W^{\pm}$ channel over $\ell^{\pm}\ell^{\pm}$
channel.

\begin{figure}[!hptb]
\centering
\includegraphics[width=11.0cm, height=8cm]{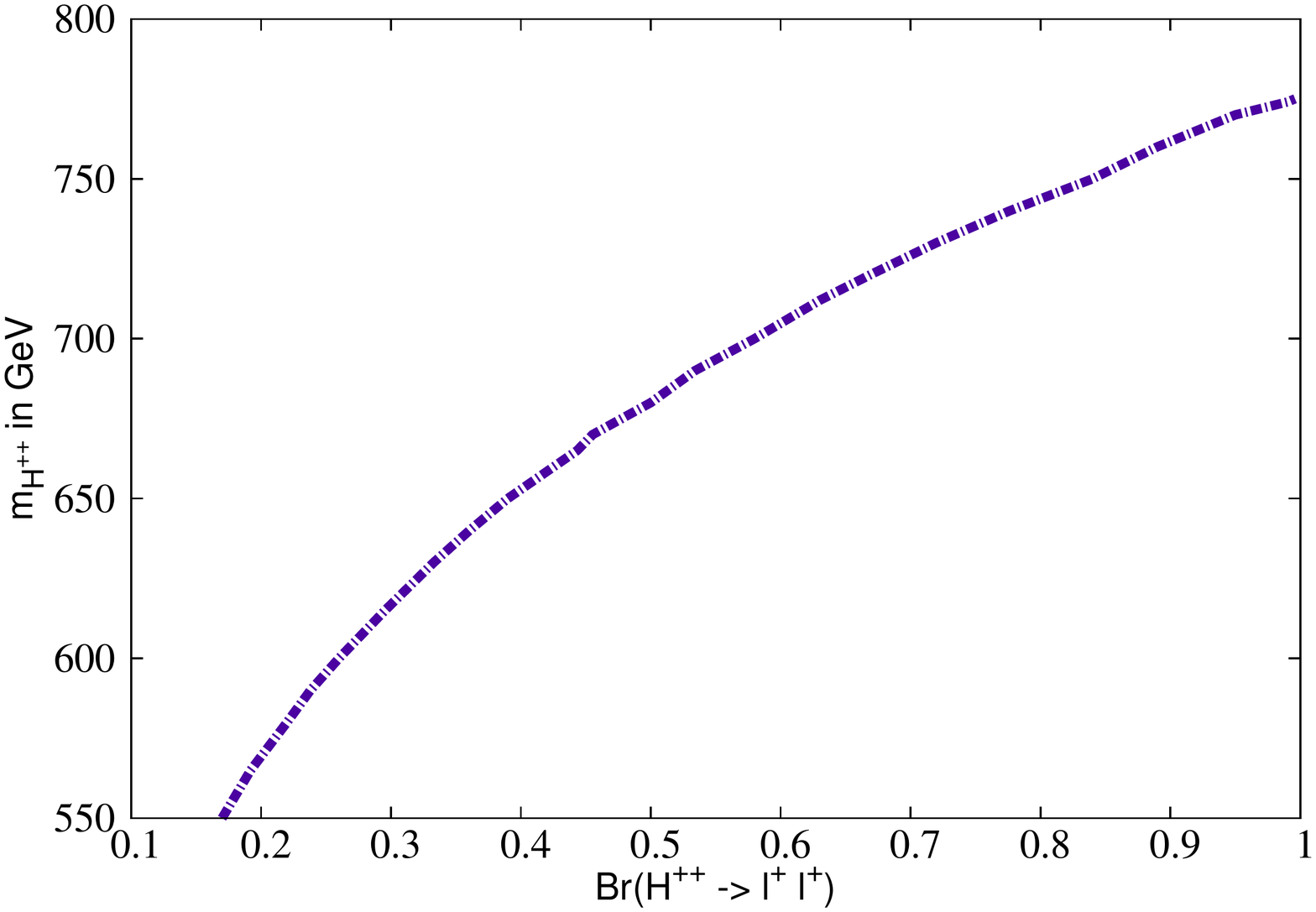}
\caption{Experimental lower limit on $m_{H^{\pm\pm}}$ as a function of Br($H^{\pm\pm} \rightarrow \ell^{\pm}\ell^{\pm}$). }
\label{brllmHpp}
\end{figure}

The doubly charged Higgs has been searched by ATLAS and CMS
collaborations. The searches focus on $H^{\pm\pm}$ produced via DY
process which is the only relevant channel. ATLAS have searched for
the DY pair production of $H^{++}H^{--}$ with 36$fb^{-1}$ data at 13
TeV in $W^{\pm} W^{\pm}$~\cite{Aaboud:2018qcu} and
$\ell^{\pm} \ell^{\pm}$~\cite{Aaboud:2017qph} channel. CMS have also looked for
$H^{\pm \pm} \rightarrow \ell^{\pm}\ell^{\pm}$ in the $H^{++}H^{--}$ and $H^{\pm \pm}H^{\mp}$
final state with 12.9 $fb^{-1}$ data at 13
TeV~\cite{CMS-PAS-HIG-16-036}. The search in the $W^{\pm}W^{\pm}$ channel puts
a lower bound of $m_H^{\pm\pm} \lsim 220$ GeV. The lower limit on
$m_{H^{\pm\pm}}$, from searches in the $\ell^{\pm}\ell^{\pm}$ final state depend on
the Br($H^{\pm\pm} \rightarrow \ell^{\pm}\ell^{\pm}$). In
Figure~\ref{brllmHpp} we show the lower limit on the mass of doubly
charged Higgs as function of Br($H^{\pm\pm} \rightarrow
\ell^{\pm}\ell^{\pm}$). One can see from this figure that the lower limit on
$m_H^{\pm\pm}$ ranges from $m_H^{\pm\pm} > 550$ GeV for Br($H^{\pm\pm}
\rightarrow \ell^{\pm}\ell^{\pm}) \simeq 17\%$ to $m_H^{\pm\pm} > 770$ GeV for
Br($H^{\pm\pm} \rightarrow \ell^{\pm}\ell^{\pm}) \simeq 100\%$.

\subsection{Same-sign dilepton channel}\label{subsec4.1}

\begin{figure}[!hptb]
\centering
\includegraphics[width=12.0cm, height=8cm]{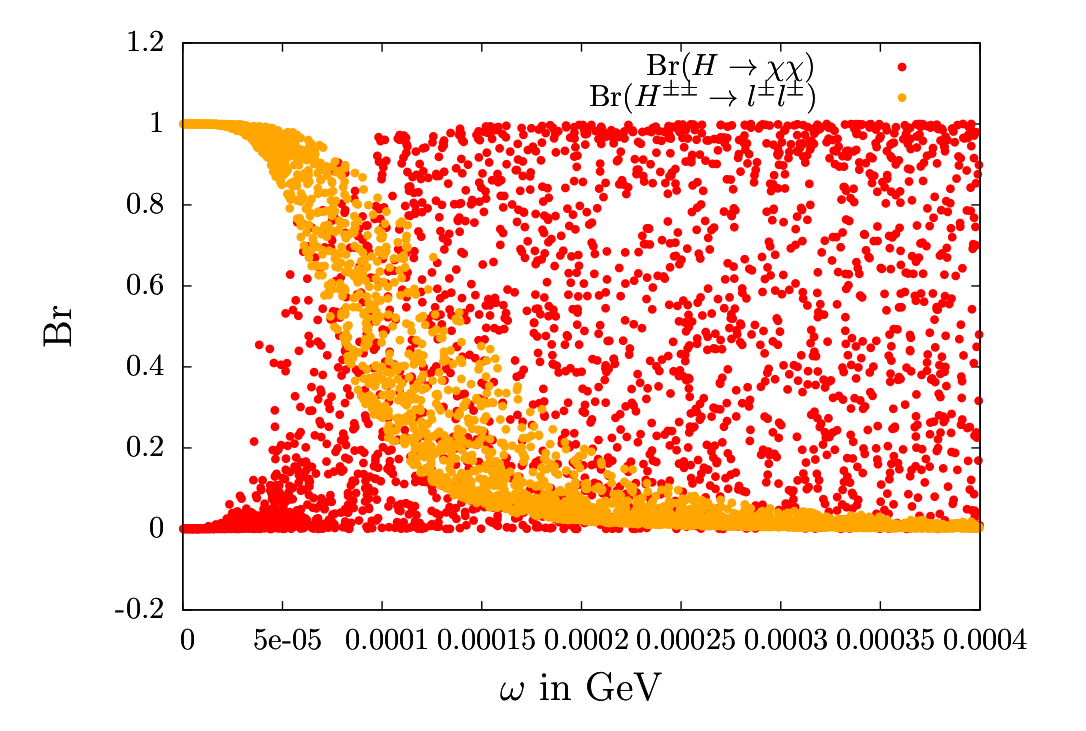}
\caption{Distribution of various branching ratios as a function of $\omega$.}
\label{omegabr}
\end{figure}

\begin{figure}[!hptb]
\centering
\includegraphics[width=8.0cm, height=11cm,angle=-90.0]{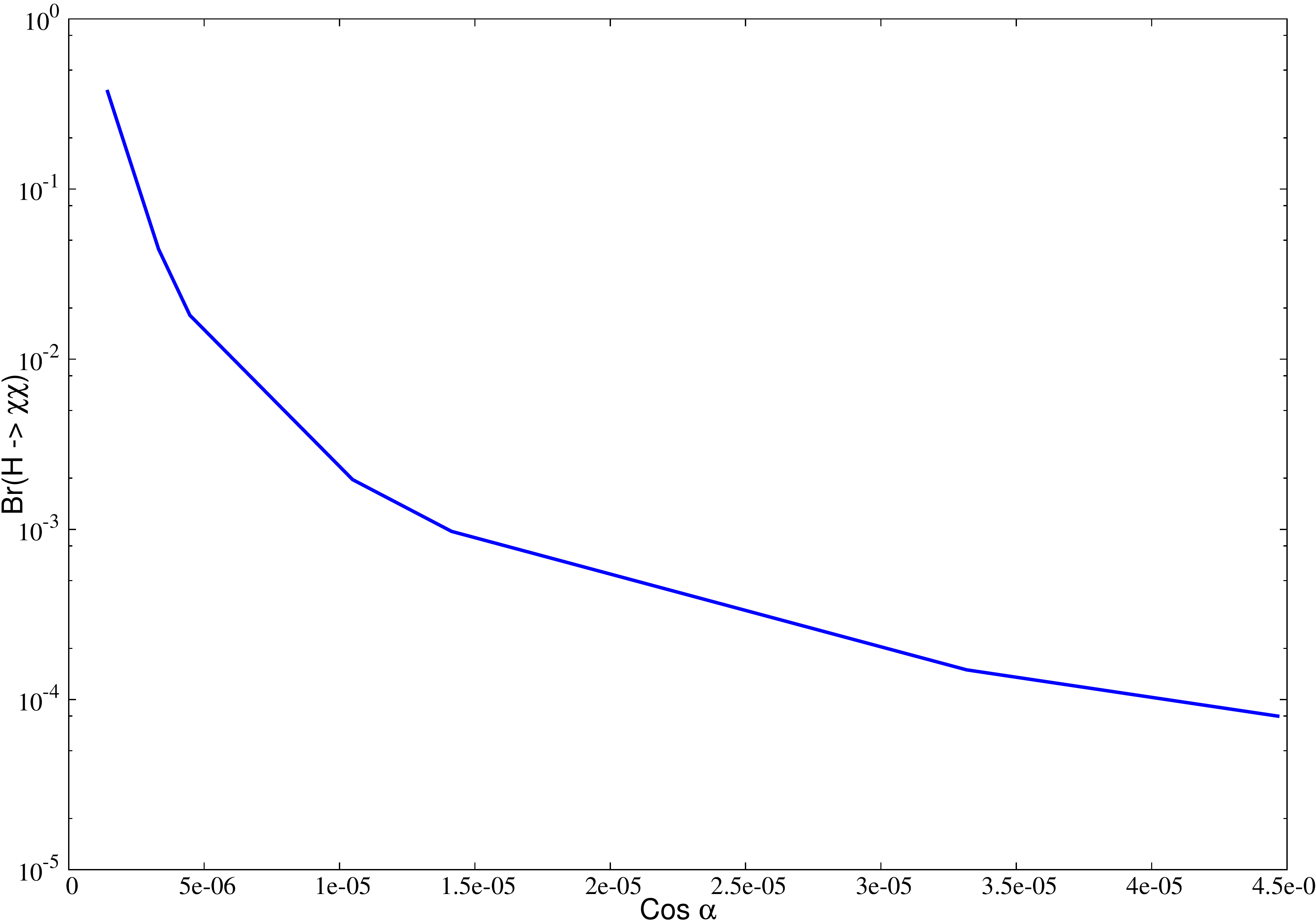}
\centering
\caption{Br($H \rightarrow \chi \chi$) as a function of the mixing angle.}
\label{calpha_br}
\end{figure}

We first look for  benchmarks for the case where
$H^{\pm\pm}$ is best looked for in the same-sign dilepton final state. We
want to identify the regions of parameter space where one can get
sizable signal events in the decay channel that we are considering. It
is clear from our previous discussion that the signal rate will be
dependent on the product of three branching ratios, namely
Br($H^{\pm\pm} \rightarrow \ell^{\pm}\ell^{\pm}$), $Br(H^\pm \rightarrow HW^{\pm})$
and Br($H \rightarrow invisible)$. It is clear from Figure~\ref{llwwbr}
that $\omega \leq$ 0.0005 GeV Br$(H^{\pm \pm} \rightarrow \ell^{\pm}
\ell^{\pm}) > 90\%$. We have noticed that when the mass gap between
$H^{\pm}$ and $H$ exceeds $m_W$, $H^{\pm}$ goes to $HW^{\pm}$ with
50\% branching as long as $\omega$ is very small. This is because, triplet
VEV and correspondingly doublet-triplet mixing being small, additional
channels such as $H^{\pm} \rightarrow hW^{\pm}$, $H^{\pm} \rightarrow
W^{\pm}Z$ and $H^{\pm} \rightarrow t \bar{b}$ do not open up. In
Figure~\ref{omegabr} we show Br($H \rightarrow \chi \chi$) as a function
of triplet VEV and also compare it with Br($H^{\pm\pm} \rightarrow
\ell^{\pm}\ell^{\pm}$). We can see that Br($H \rightarrow \chi \chi$)
increases with increasing triplet VEV whereas $Br(H^{\pm\pm}
\rightarrow \ell^{\pm}\ell^{\pm}$) decreases with it. Typically one can
choose some intermediate $\omega \in [10^{-5}, 10^{-4}]$ to get
moderately good branching ratios in both these channels at the same
time. We also notice that unless the mixing between the doublet and
triplet CP-even scalar states is extremely small, the $H$ goes
primarily to a pair of $hh$ and consequently Br($H \rightarrow \chi \chi$)
becomes very small. The dependence of Br($H \rightarrow \chi \chi$) on the mixing angle $\alpha$ is
shown in Figure~\ref{calpha_br}. Therefore to get considerable branching
in the $H \rightarrow \chi \chi$ channel, we have taken the mixing to
be very small, ie. $\sin \alpha \sim 1$.

One should be careful while calculating the invisible decay width of
heavy Higgs in this case, since $H$ can go to a pair of neutrinos or
antineutrinos when the lepton flavor violating yukawa coupling is
large enough. That will also contribute to invisible decay of the
heavy Higgs. Br($H \rightarrow \nu \nu/\bar{\nu} \bar{\nu}$) has same dependence
on $\omega$ as Br($H^{\pm\pm}
\rightarrow \ell^{\pm}\ell^{\pm}$), because they are governed by the same
yukawa coupling. We will consider Br$_{invisible}$ of heavy Higgs to
be the sum of Br($H \rightarrow \chi \chi$) and Br($H \rightarrow \nu
\nu/\bar{\nu} \bar{\nu}$). We have chosen our benchmark points in a
way to encompass different scenarios. We have chosen two cases (BP1
and BP2). In BP 1 Br($H \rightarrow \nu \nu/\bar{\nu} \bar{\nu}$)
dominates over Br($H \rightarrow \chi \chi$), and in BP 2 they are
comparable and we have tried to see whether these two cases can be
distinguished. For comparison we
have kept $m_H$ in a similar region in the two cases. We
choose a third benchmark (BP 3) with lower $m_H$ and chosen
$\omega$ in such a way that Br($H \rightarrow \chi \chi$) dominates
over Br($H \rightarrow \nu \nu/\bar{\nu} \bar{\nu}$). In this case
although the total branching in the specific decay mode will be less,
the low mass of $H$ will enable us to get larger production cross
section and in turn can be probed at the LHC.

\begin{figure}[!hptb]

\includegraphics[width=12.0cm, height=10cm]{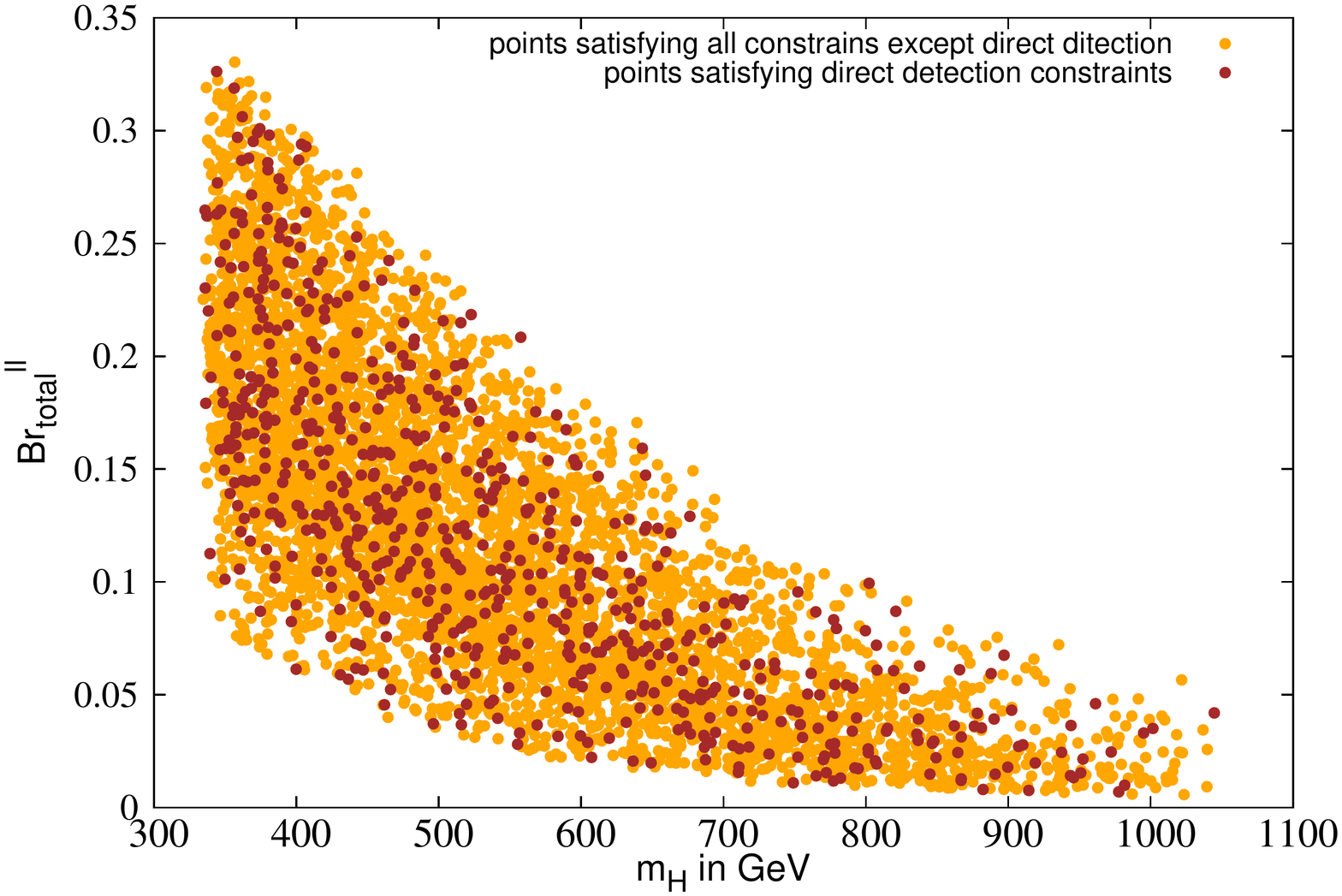}
\centering
\caption{$m_{H}$ distribution of $Br_{total}^{\ell\ell}$. Orange points
  satisfy only relic density bounds and over and above that the brown
  points satisfy the current direct detection limit coming from
  XENON1T.}
\label{brtotmH}
\end{figure}

We define a new parameter $Br_{total}^{\ell\ell} = Br(H^{\pm \pm} \rightarrow
\ell^{\pm} \ell^{\pm}) \times Br(H^{\pm} \rightarrow W^{\pm} H) \times Br(H
\rightarrow invisible)$ and search for moderate to large values of
this quantity throughout our allowed parameter space. In
Figure~\ref{brtotmH} we plot $Br_{total}^{\ell \ell}$ as a function of $m_H$. The
 orange region satisfy all the constraints except direct
detection. The brown points satisfy the direct detection constraints
along with all other constraints discussed above. We present our benchmark choices governed by the discussion above in
Table~\ref{benchmark}. We have checked that they obey all the
constraints discussed in Section~\ref{sec3}, including the relic density suggested by the Planck data at $2\sigma$ level..\\

\begin{table}[!hptb]
\begin{center}
\begin{footnotesize}
\begin{tabular}{| c | c | c | c |}
\hline
& BP 1 & BP 2 & BP 3\\
\hline
$m_H$ in GeV & 423.1 & 615.1 & 615.1\\
\hline
$m_A$ in GeV & 423.1 & 615.1 & 615.1 \\
\hline
$m_{H^{\pm}}$ in GeV  & 509.3 & 697.0 & 697.0 \\
\hline
$m_{H^{\pm \pm}}$ in GeV  & 582.8 & 770.0 & 770.0 \\
\hline
$m_\chi$ in GeV  & 59.3 & 56.4 & 56.4\\
\hline
$\lambda_S$ & 0.49 & 0.0297 & -0.0297\\
\hline
$\lambda_D$ & 0.00069 & 0.002125 & 0.002125\\
\hline
$\lambda_T$ & 11.258 & 10.51 & 10.51\\
\hline
$\omega$ in GeV  & 1.348$\times 10^{-4}$ & 4.074$\times 10^{-5}$ & 7.274$\times 10^{-5}$\\
\hline
$\sigma(p p \rightarrow H^{\pm \pm} H^{\mp})$ in $fb$ & 1.19 & 0.43 & 0.43\\
\hline
$Br(H \rightarrow invisible)$ & 0.92 & 0.935 & 0.79\\
\hline
$Br(H^{\pm \pm} \rightarrow \ell^{\pm} \ell^{\pm})$ & 0.228 & 0.95 & 0.65\\
\hline
$Br_{total}$ & 0.1049 & 0.44365 & 0.25675 \\
\hline
\end{tabular}
\end{footnotesize}
\caption{The Benchmark points for same-sign dilepton channel.}
\label{benchmark}
\end{center}
\end{table}

\subsection{Same-sign vector boson ($W^{\pm}W^{\pm}$) channel}\label{subsec4.1} 

\begin{figure}[!hptb]
\centering
\includegraphics[width=12.0cm, height=8cm]{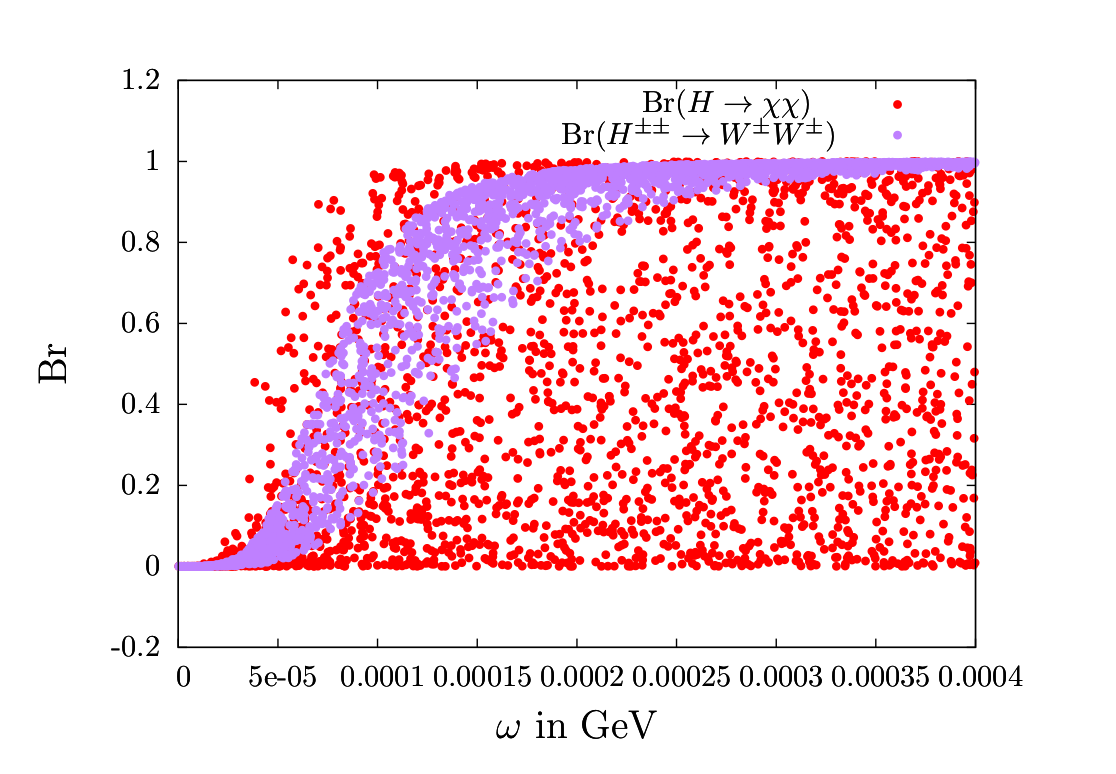}
\caption{Distribution of various branching ratios as a function of triplet VEV.}
\label{omegabrww}
\end{figure}

We turn next to the other important decay mode of $H^{\pm\pm}$, namely, a
pair of same-sign $W$ bosons, which will give rise to different
signature. In
Figure~\ref{omegabrww} we present the comparison between $Br(H^{\pm\pm} \rightarrow
W^{\pm}W^{\pm}$) and $Br(H^{\pm\pm} \rightarrow \chi\chi$), the two relevant
branching fractions in this case. We can see
here that Br$(H^{\pm\pm} \rightarrow W^{\pm}W^{\pm})$ increases with $\omega$ and becomes nearly 100\% for $\omega \gsim
10^{-4}$ GeV. This is because when the triplet VEV increases beyond this value, $Br(H^{\pm\pm} \rightarrow \ell^{\pm}\ell^{\pm})$ becomes very low
due to suppression in the lepton number violating Yukawa coupling
and therefore the $W^{\pm}W^{\pm}$ channel takes over. As a consequence of the concomitantly
suppressed lepton number violating Yukawa coupling $Br(H
\rightarrow \nu\nu/\bar{\nu}\bar{\nu})$ also decreases significantly
and therefore the heavy Higgs dominantly goes into the $\chi\chi$ channel. Thus in Figure~\ref{omegabrww} both $Br(H^{\pm\pm} \rightarrow
W^{\pm}W^{\pm})$ and $Br(H^{\pm\pm} \rightarrow \chi\chi)$ both
increase as $\omega$ increases. A notable point here is that in this
region with larger triplet VEV, the invisible branching ratio of $H$ will
consist of $H \rightarrow \chi \chi$ channel overwhelmingly, because of
negligible branching fraction of $H$ in the
$\nu\nu/\bar{\nu}\bar{\nu}$ channel.

  While choosing benchmarks for our collider analysis we keep in mind the extremely low leptonic branching ratio of the same-sign $W$ pair. Therefore to get sufficient event rate we have chosen mass of $H$ to be on the lower side (220-400 GeV) which are consistent with the experimental searches. In BP 1 $m_H$ has been chosen to be $\simeq 220$ GeV. In BP 2 and BP 3 we take $m_H$ in a slightly higher range around $ 300-400$ GeV. When the triplet VEV is small and correspondingly the doublet-triplet mixing is also low, the decay modes $H^{\pm} \rightarrow hW^{\pm}$, $H^{\pm} \rightarrow W^{\pm}Z$ and $H^{\pm} \rightarrow t \bar{b}$ are not accessible. Hence $Br(H^{\pm} \rightarrow HW^{\pm}$) and $Br(H^{\pm} \rightarrow AW^{\pm}$) become the two dominant decay channels, each about 50\% branching ratio as was discussed in the previous subsection. But as the triplet VEV increases, doublet-triplet mixing also goes up and the modes $H^{\pm} \rightarrow hW^{\pm}$, $H^{\pm} \rightarrow W^{\pm}Z$ and $H^{\pm} \rightarrow t \bar{b}$ open

\begin{figure}[!hptb]

\includegraphics[width=12.0cm, height=10cm]{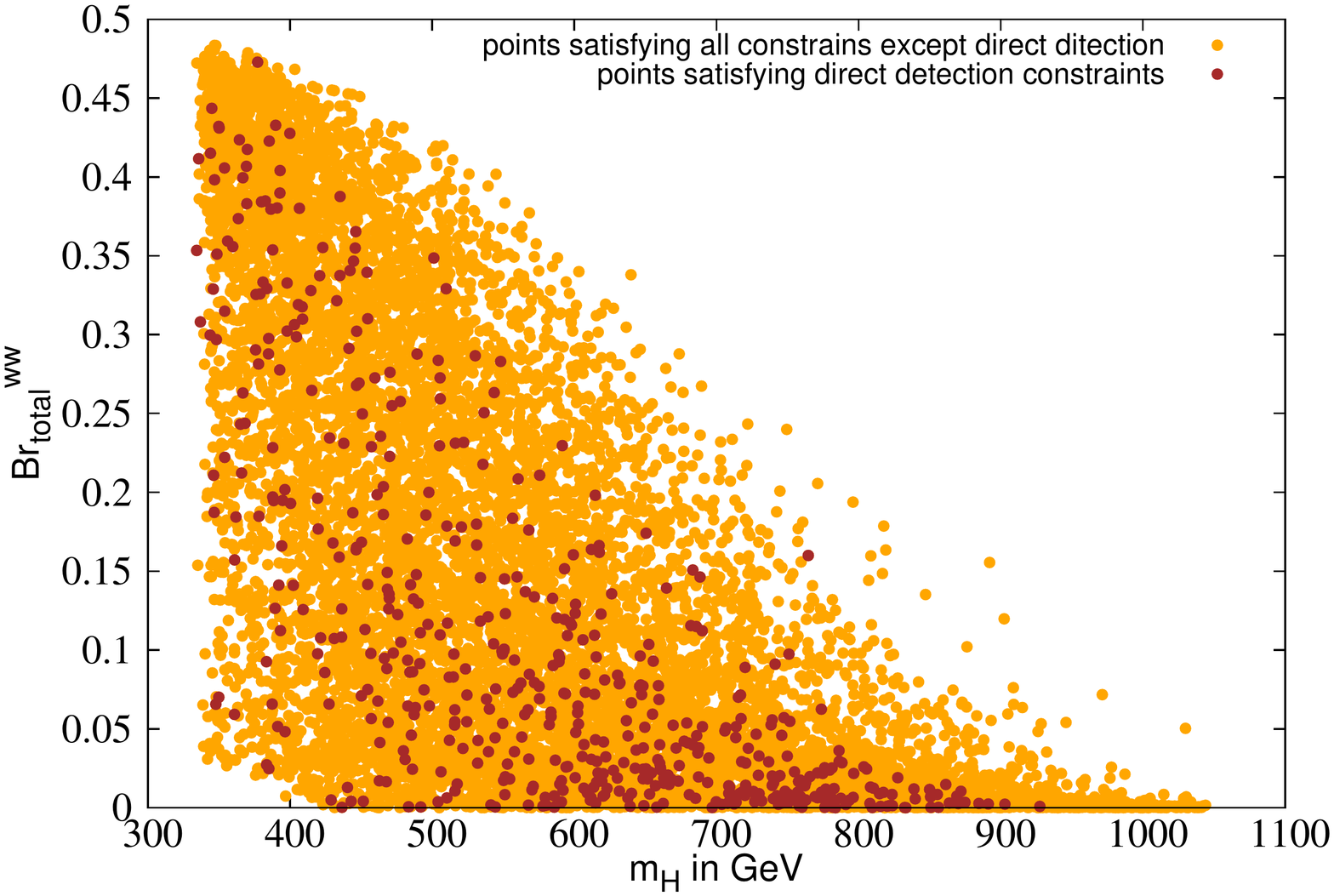}
\centering
\caption{$m_{H}$ distribution of Br$_{total}^{WW}$. Orange points satisfy only relic density bounds and over and above that the brown points satisfy the current direct detection limit coming from XENON1T.}
\label{brtotmHww}
\end{figure}

 \begin{table}[!hptb]
\begin{center}
\begin{footnotesize}
\begin{tabular}{| c | c | c | c |}
\hline
& BP 1 & BP 2 & BP 3\\
\hline
$m_H$ in GeV & 220.0 & 300.0 & 400.0\\
\hline
$m_A$ in GeV & 220.0 & 300.0 & 400.0 \\
\hline
$m_{H^{\pm}}$ in GeV  & 301.0 & 382.0 & 482.0 \\
\hline
$m_{H^{\pm \pm}}$ in GeV  & 371.0 & 451.0 & 551.0 \\
\hline
$m_\chi$ in GeV  & 57.6 & 125.0 & 180.1 \\
\hline
$\lambda_S$ & 0.0472 & 0.0725 & 0.0264\\
\hline
$\lambda_D$ & 0.00156 & 0.00862 & 0.0256\\
\hline
$\lambda_T$ & 8.67 & 5.3938 & 8.981\\
\hline
$\omega$ in GeV  & 0.1034 & 4.68 & 0.261\\
\hline
$\sigma(p p \rightarrow H^{\pm \pm} H^{\mp})$ in $fb$ & 16.2 & 6.7 & 2.5 \\
\hline
$Br(H \rightarrow invisible)$ & 0.99 & 0.89 & 0.82 \\
\hline
$Br(H^{\pm} \rightarrow W^{\pm} H)$ & 0.50 & 0.30 & 0.50 \\
\hline
$Br(H^{\pm \pm} \rightarrow W^{\pm} W^{\pm})$ & 1.0 & 1.0 & 1.0\\
\hline
$Br_{total}$ & 0.49 & 0.27 & 0.41 \\
\hline
\end{tabular}
\end{footnotesize}
\caption{The Benchmark points for same-sign $W$ channel.}
\label{benchmark2}
\end{center}
\end{table}

\noindent
up with considerable branching fractions. Consequently, $Br(H^{\pm} \rightarrow HW^{\pm}$) falls. In BP 2 we have considered such a situation with $\omega$ close to its allowed upper limit. In this case $Br(H^{\pm} \rightarrow HW^{\pm})$  comes down to 30\% (see Table~\ref{benchmark2}).

In Figure~\ref{brtotmHww} we plot the quantity $Br_{total}^{WW}$ analogous to $B_{total}^{\ell\ell}$ as defined in the previous subsection, as a function of $m_H$ when the $W^\pm W^\pm$ decay mode of the doubly charged Higgs becomes dominant. The three benchmark points, used in our study of the $W^\pm W^\pm$-driven final state, are shown in Table~\ref{benchmark2}. Once more, these are consistent with all constraints including those from the observed relic density.

\section{Collider Analysis (Cut based)}\label{sec5}

From the discussion of the previous section, we are convinced that the heavy neutral Higgs can serve as a dark matter portal in  a Type-II Seesaw scenario with a singlet scalar DM particle. Our goal at this point is to look for signatures of this model in the channels already discussed in the previous section, and explore their reach at the high-luminosity run of the LHC. In this spirit, we consider in turn cases where the heavy CP-even Higgs ($H$) can decay into a pair of dark matter with substantial branching fraction. Obviously, the events will consist of large $\slashed{E_T}$. As mentioned already, production of $H$ can be significant only through Drell-Yan processes. Hence we concentrate on 
(i)  $p p \rightarrow H^{\pm\pm} H^{\mp}, H^{\pm\pm} \rightarrow \ell^{\pm}\ell^{\pm}, H^{\pm} \rightarrow H W^{\pm}, H \rightarrow invisible$, and (ii) $p p \rightarrow H^{\pm\pm} H^{\mp}, H^{\pm\pm} \rightarrow W^{\pm}W^{\pm}, H^{\pm} \rightarrow H W^{\pm}, H \rightarrow invisible$.
These two channels are somewhat complementary in nature, having significant rates in  
different regions of the parameter space. We will henceforth call the first scenario Case I, 
and second one, Case II. As has been stated in the introduction, we have also considered the $W$-boson fusion process, namely, $p p \rightarrow H^{\pm\pm} H + \text{two forward jets}$ after which $H$ decays into invisible final states. However, this process will have irreducible background from SM VBF production and will not have enough signal rate even at the high-luminosity(HL) LHC. 
Thus we will concentrate on the DY-production of $H$ with final states pertaining to the two major decay modes of  $H^{\pm\pm}$, namely, $\ell^{\pm}\ell^{\pm}$ and $W^{\pm}W^{\pm}$. We will
briefly comment on the $W$-fusion channel  at the end of this section.   

Events for the signals and their corresponding backgrounds have been generated using Madgraph@MCNLO~\cite{Alwall:2014hca} and their cross-sections have been calculated at the next-to-leading order(NLO). We take the renormalization and factorization scales at the $p_T$ of the hardest jet and also use the nn23lo1 parton distribution function. At the NLO level, the results 
with other scale choices do not differ by more than ($10\%$). PYTHIA8~\cite{Sjostrand:2006za} has been used for the showering and hadronization and the detector simulation has been taken care of by Delphes-3.4.1~\cite{deFavereau:2013fsa}.

\subsection{Case I}\label{sub5.1}

The Drell-Yan production of $H^{\pm\pm}H^{\mp}$ will lead to the final state containing a pair of same-sign dilepton from the decay of $H^{\pm\pm}$. The $H^{\pm}$ will decay into $W^{\pm}$ and $H$ wherever this decay is kinematically allowed \footnote{Beyond the kinematic limit for this two-body 
decay, while the $l \nu$ channels become appreciable, the decay products of $W^{\pm} H$ 
still dominate the final state so long as the level of `off-shellness' is not too high, as happens in the regions of our interest.}. The invisible decay of $H$ will lead to 
$\slashed{E_T}$ in the final state. We have considered only hadronic decays of $W^{\pm}$ to have sizable number of events in the signal process. The same-sign dilepton pair constitutes
 a clean signal to look for in experiments. 

\medskip

\noindent
{\bf Signal:} The signal here is a pair of same-sign leptons ($e/\mu$) + 2 jets + $\slashed{E_T}$. This signal has been searched for in the LHC~\cite{Khachatryan:2016kod}. It reports no significant excess over the SM expectation with $\int{\cal L} dt = 36 fb^{-1}$ at 95\% C.L. .

\medskip

\noindent
{\bf Background:} The dominant backgrounds for this final state are~\cite{Khachatryan:2016kod} 

\begin{itemize}
\item $t \bar t$ semileptonic decay which leads to non-prompt leptons in the final state. Non-prompt leptons are those which can arise from heavy flavor decay or hadrons being misidentified as leptons etc. 
\item $W$ + jets also contributes to the background producing non-prompt leptons. 
\item $t \bar t W^{\pm}$ with semileptonic decay of $t \bar t$ which directly produces same-sign dilepton background is another background. 
\item $W^{\pm}Z$ with leptonic decay of $W^{\pm}$ and $Z$ also produces same-sign dilepton pairs and therefore is an important background for our signal.
\item  Charge misidentification: The charge misidentification probability for $e^{\pm}e^{\pm}$ lies in the range $10^{-5} - 10^{-3}$~\cite{Khachatryan:2016kod} depending on the $p_T$ and $\eta$. For muons charge misidentification probability is negligible~\cite{Khachatryan:2016kod}. This background thus does not play any significant role in the analysis. 
\end{itemize}

\subsubsection{Distributions}

\begin{figure}[!hptb]

\includegraphics[width=8.8cm, height=7cm]{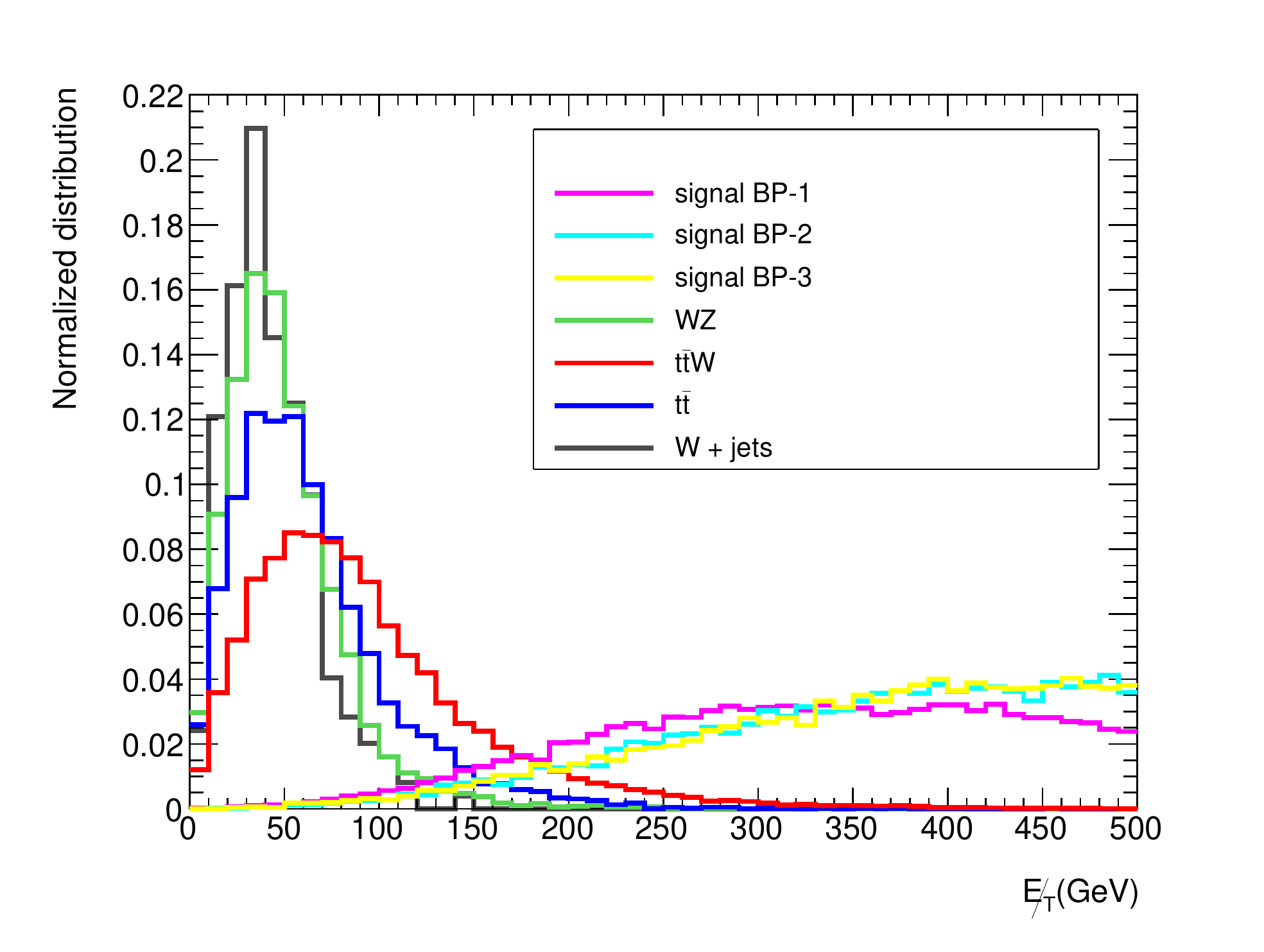}
\includegraphics[width=8.8cm, height=7cm]{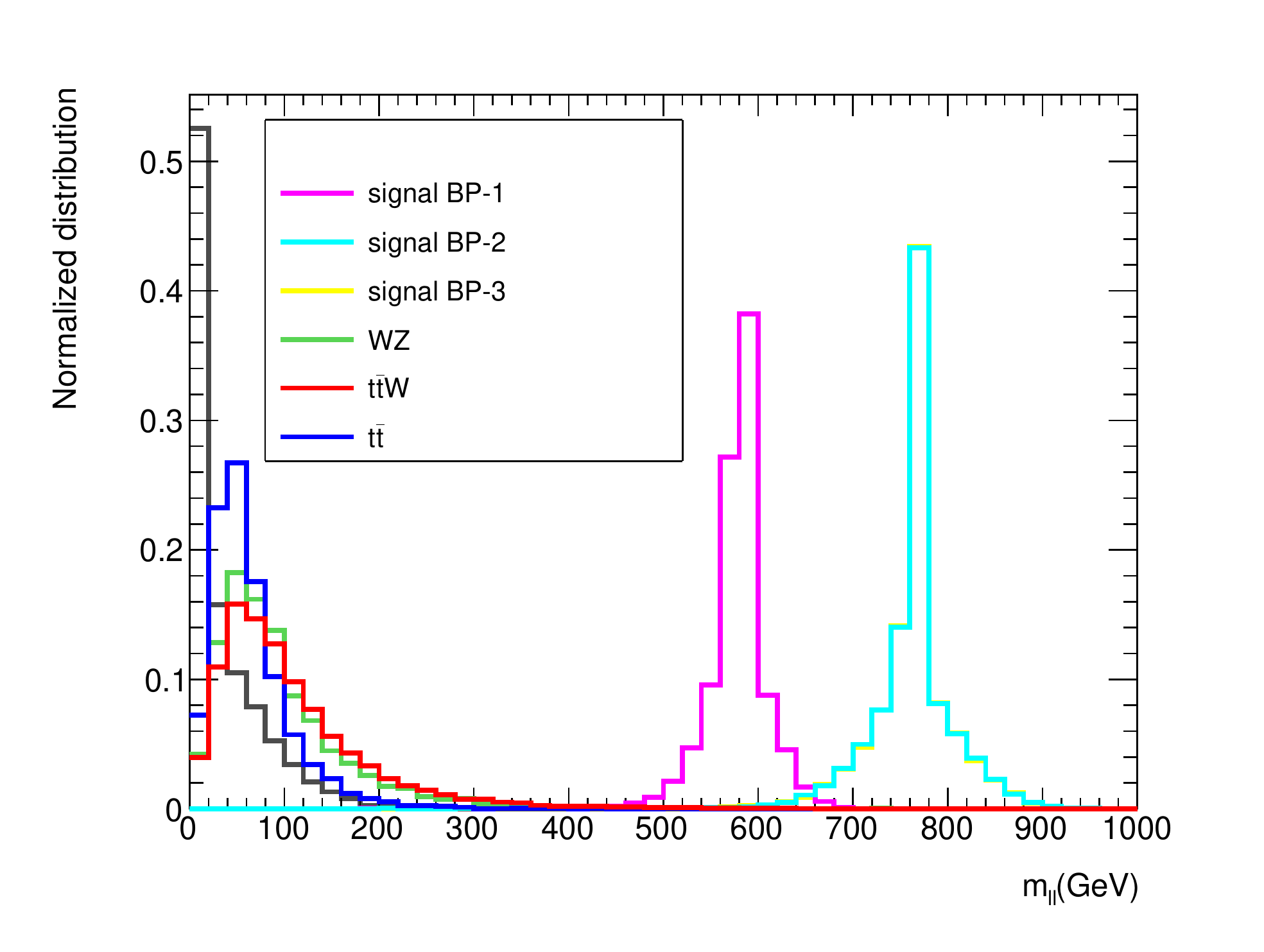}

\caption{Distribution of $\slashed{E_T}$(left) and invariant mass(right) of same-sign dileptons for the three signal BPs and backgrounds 
in case I.}
\label{misspt_invll_ll}
\end{figure}

\begin{figure}[!hptb]

\includegraphics[width=8.8cm, height=7cm]{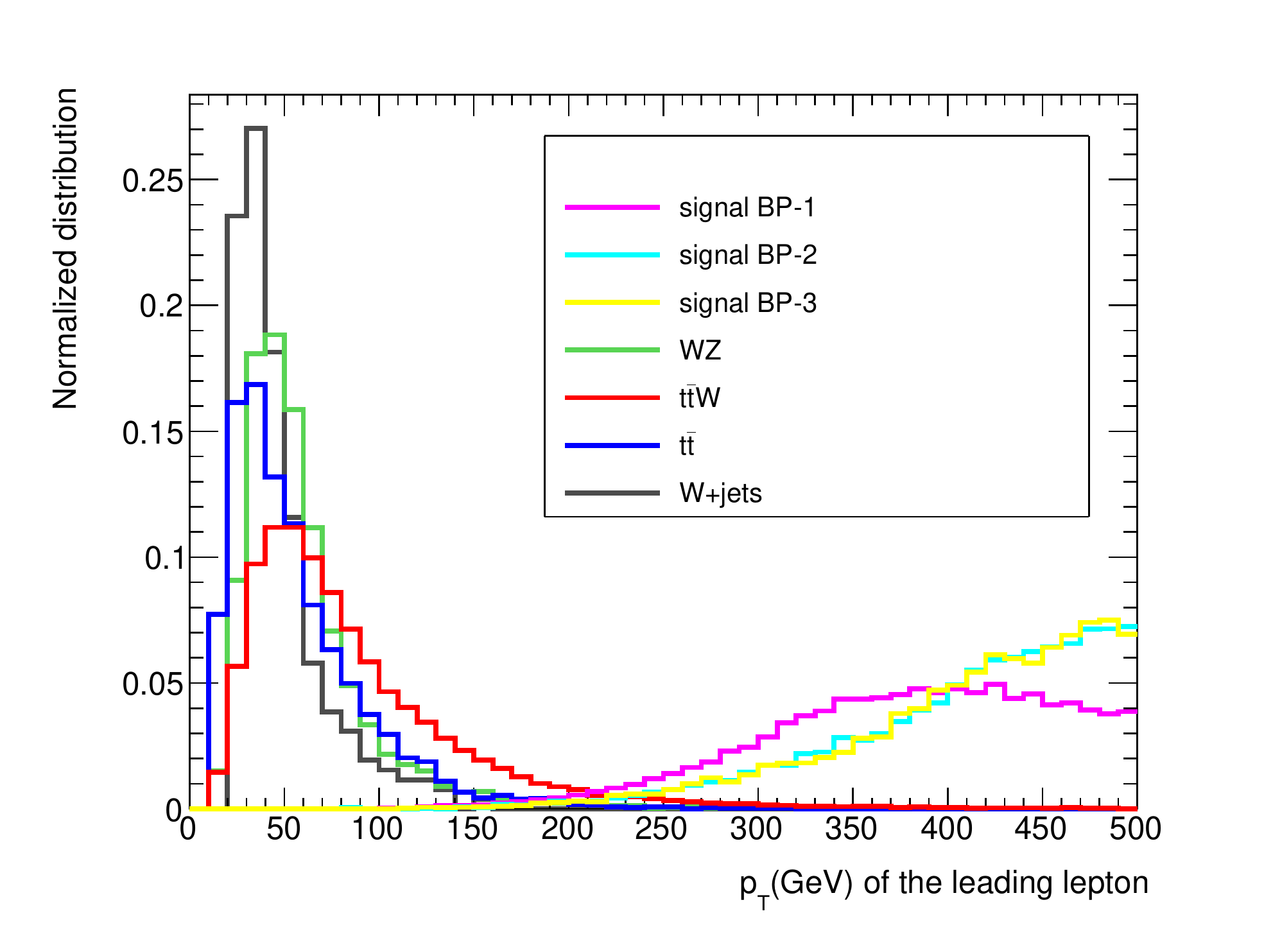}
\includegraphics[width=8.8cm, height=7cm]{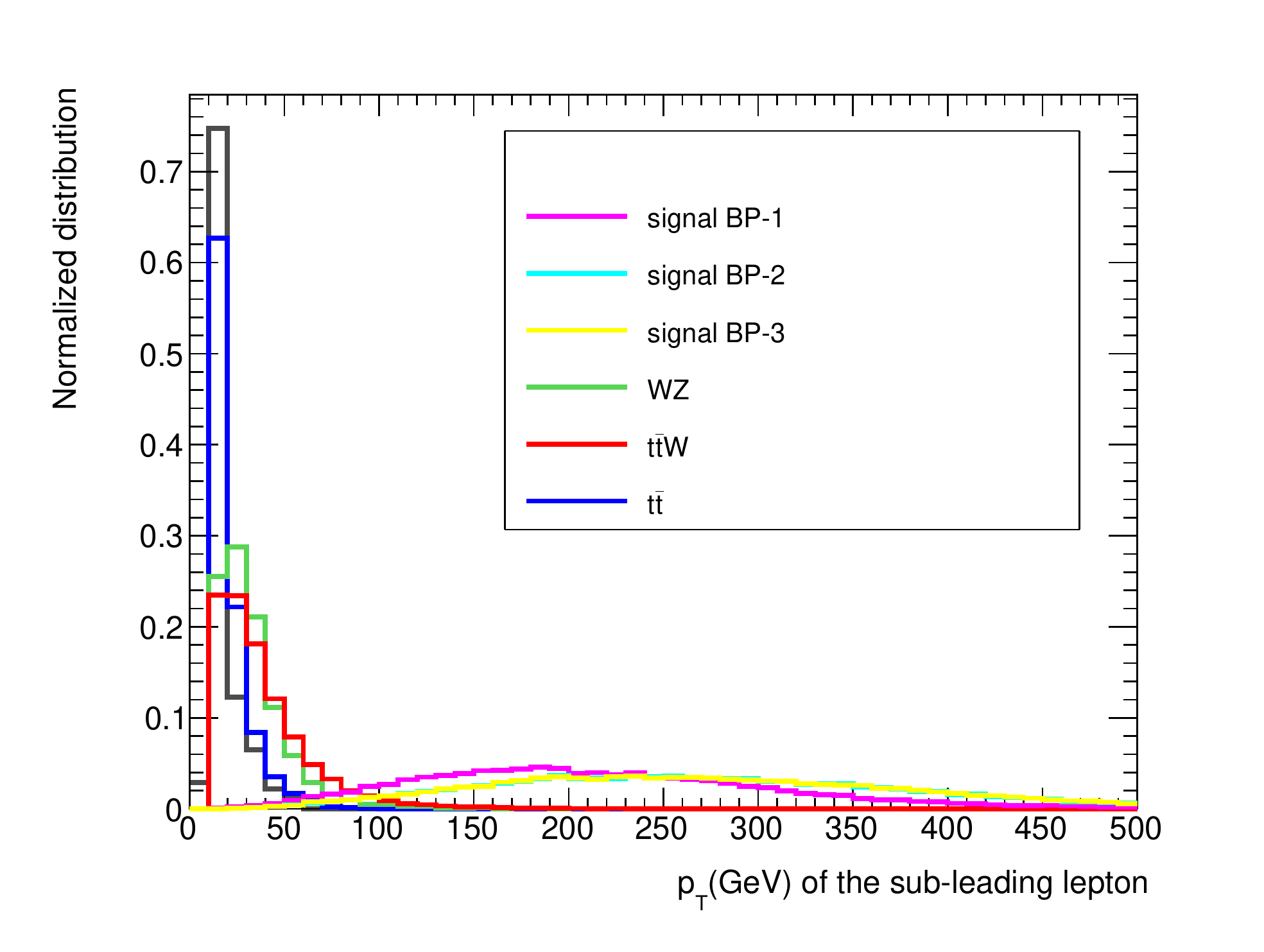}

\caption{Distribution of transverse momenta of the leading(left) and sub-leading(right) leptons for the three signal BPs and backgrounds in case I.}
\label{ptl1_ptl2_ll}
\end{figure}

\begin{figure}[!hptb]

\includegraphics[width=8.8cm, height=7cm]{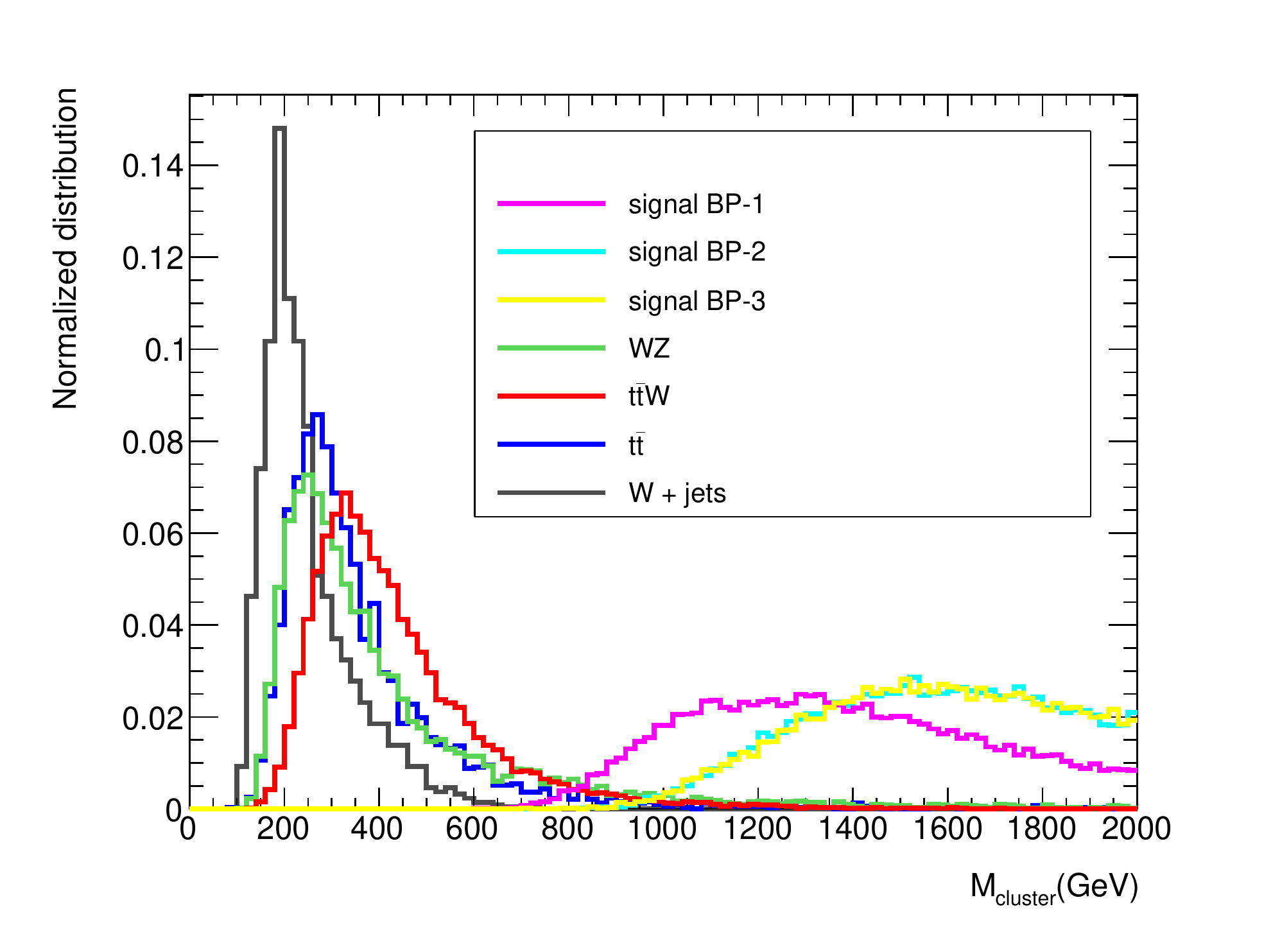}
\includegraphics[width=8.8cm, height=7cm]{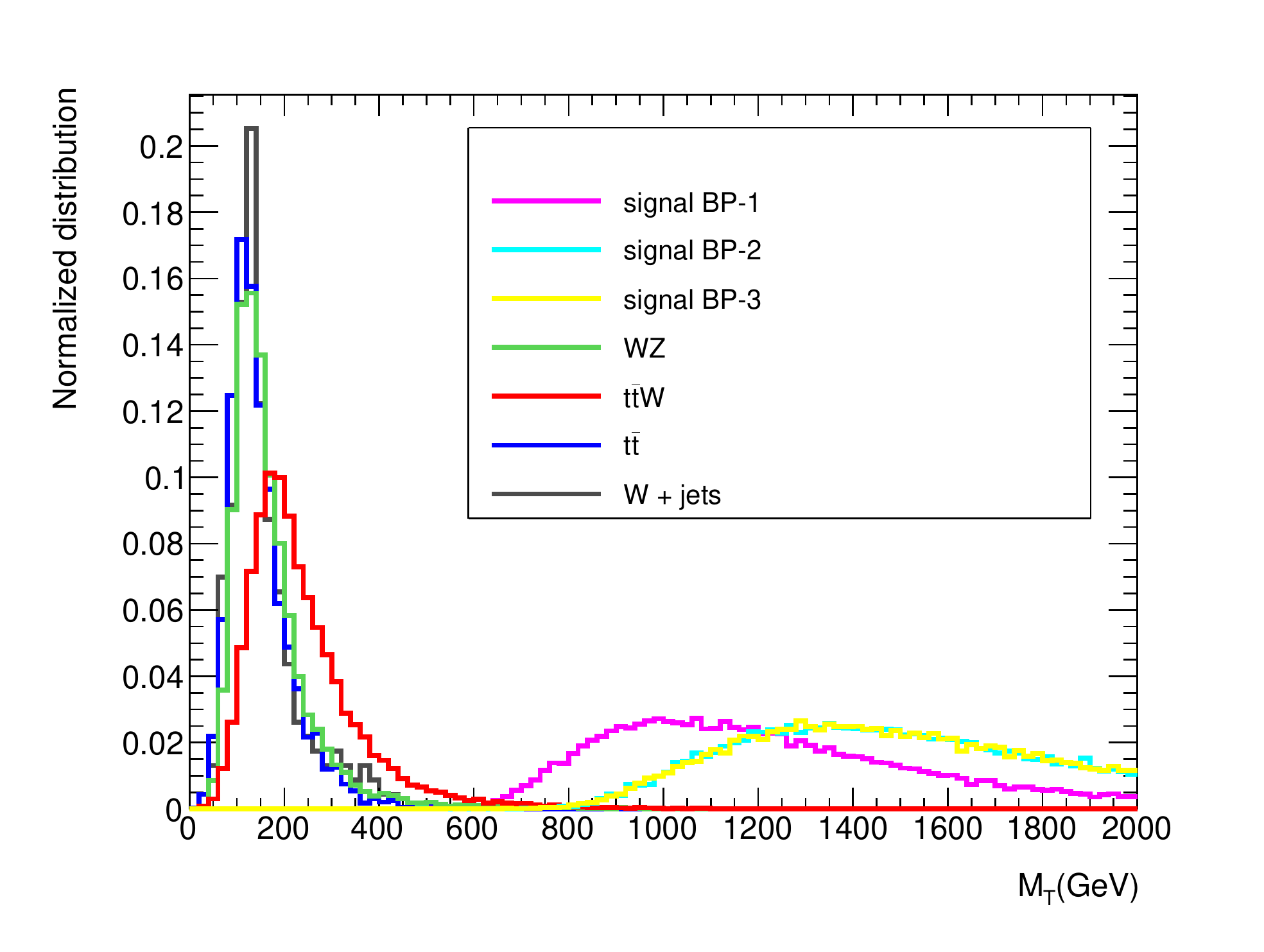}

\caption{Distribution of cluster transverse mass(left) and transverse mass(right) for the three signal BPs and backgrounds in case I.}
\label{mcl_mtr_ll}
\end{figure}

\begin{figure}[!hptb]

\includegraphics[width=8.8cm, height=7cm]{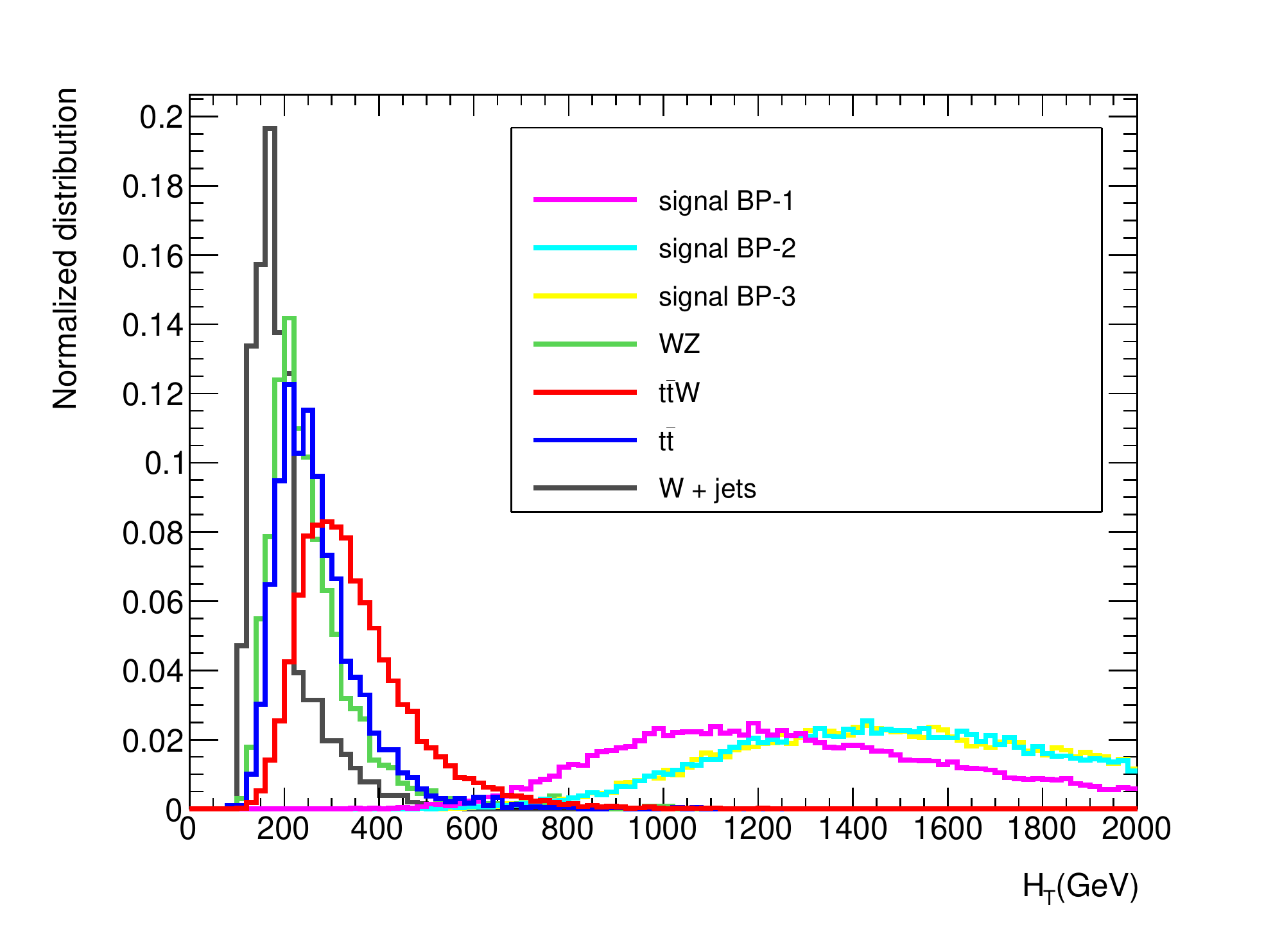}
\includegraphics[width=8.8cm, height=7cm]{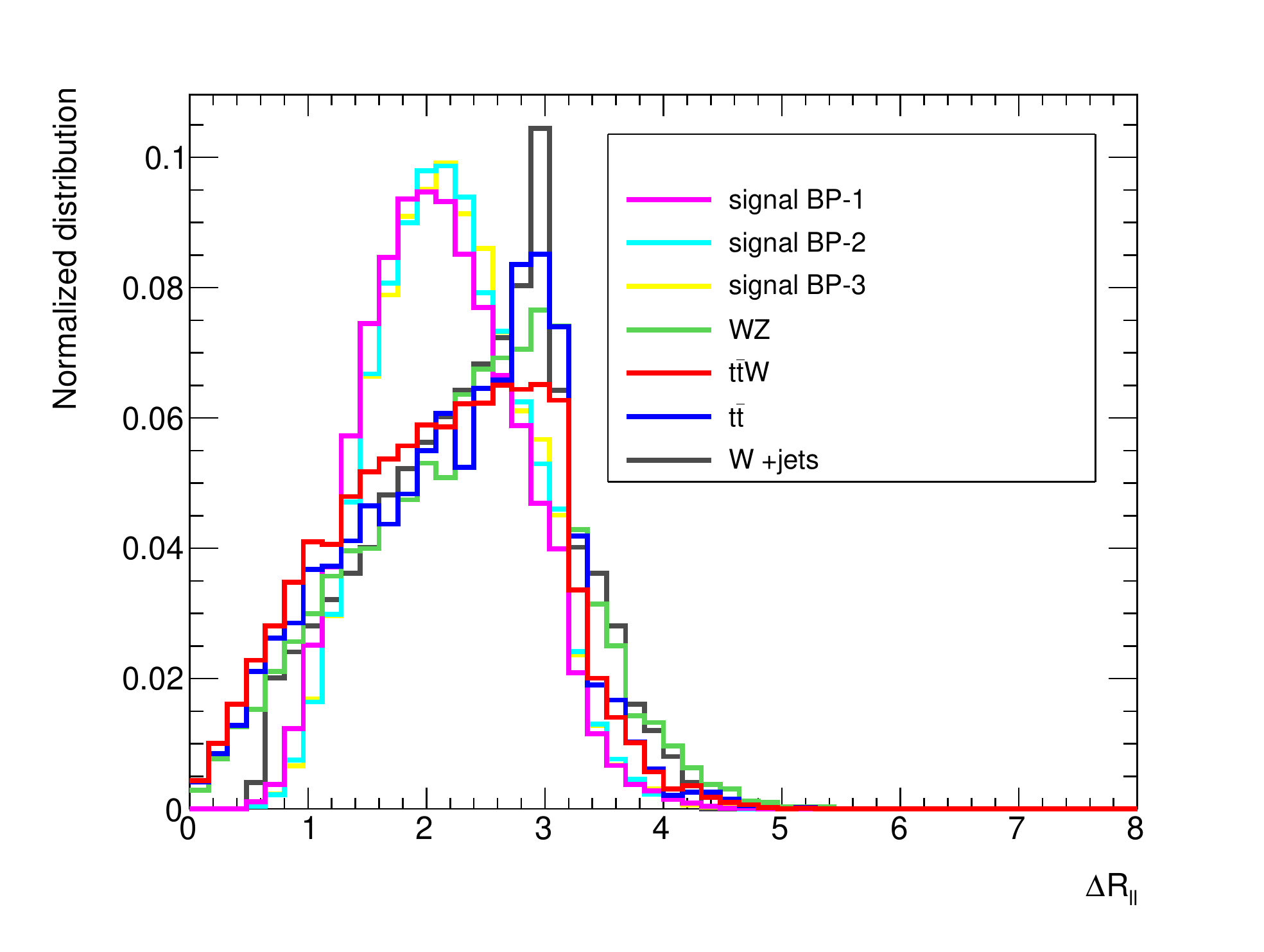}

\caption{Distribution of $H_T$ and $\Delta R$ between leading and sub-leading leptons for the three signal BPs and backgrounds in case I.}
\label{ht_drll_ll}
\end{figure}

We present various kinematic distributions for the signal and background processes. In Figure~\ref{misspt_invll_ll} (left), we plot the $\slashed{E_T}$ and invariant mass of the same-sign dilepton pair. The $\slashed{E_T}$ in the signals peaks at a higher value than that of the backgrounds since the $\slashed{E_T}$ in the signal comes from the invisible decay of a heavy Higgs. For BP 2 and 3 the $\slashed{E_T}$ peaks at a higher value as compared to BP 1, because of the higher mass of $H$ in the former case. The fact that  the invariant mass of the same-sign dilepton peaks at $m_{H^{\pm\pm}}$ adds to the distinctness of the events, as can be seen in Figure~\ref{misspt_invll_ll} (right). 
 
In Figure~\ref{ptl1_ptl2_ll} we present the $p_T$ distributions of the leading and sub-leading leptons of the same-sign dilepton pair. The $p_T$ of the leptons in case of signal is much larger than that of the backgrounds as the dilepton pair in the signal process comes from the decay of a heavy doubly-charged Higgs. These, along with the  observables mentioned in the previous paragraph, 
serve well to discriminate the signal from backgrounds.

Next come three observables which are related to each other. They are cluster transverse mass ($M_{cluster}$), transverse mass ($M_T$) and scalar $p_T$ sum ($H_T$), being defined as~\cite{Han:2007bk}

\begin{equation}
M_{cluster}  =  \sqrt{m^2_{2j} + (\sum {\vec{p_T}^j})^2}  +
 \sqrt{m^2_{\ell\ell} + (\sum{\vec{p_T}^\ell})^2}  + \slashed{E_T}
\label{mct}
\end{equation}

\begin{equation}
\label{mt}
M_T  = \sqrt{ (\sqrt{m^2_{\ell\ell}  + (\sum {\vec{p_T}^\ell})^2} + \slashed{E_T})^2}  - 
  (\sum{\vec{p_T}^\ell} + \vec{\slashed{E_T})^2}.
\end{equation}

and 

\begin{equation}
H_T  = (\sum {{p_T}^j}) + (\sum {{p_T}^\ell}) + \slashed{E_T}
\label{ht}
\end{equation}
\noindent
respectively.

From  Equations.~\ref{mct},~\ref{mt} and ~\ref{ht} we can see that $m_{cluster}$ represents the 
sum of $p_T$ of the dilepton and jets system, invariant mass of the dilepton and the jets system 
and $\slashed{E_T}$. $M_T$ represents the sum of $p_T$ of the dilepton system, invariant mass of the dilepton system and $\slashed{E_T}$. $H_T$, on the other hand is the scalar sum of the transverse momenta of all the final state particles. As Table~\ref{llcuts} shows, cuts on  these variables have practically
the same efficiency as far as the signal is concerned, while they affect the background a little
differently from each other. While they have been applied in succession in the cut-based analysis reported here, they have been retained in the subsequent neural network analyses too, where their
correlation is duly taken into account.

From Figure~\ref{mcl_mtr_ll} (left) it can be seen that the distribution in the cluster transverse mass for the whole system for the signal peaks at a higher value than that of the background. 
The $M_T$-distribution in the right panel shows a similar trend. 
Figure~\ref{ht_drll_ll} (left) shows the $H_T$-distributions, once more with the same trend, as expected. This common feature of all three observables is there because of higher $p_T$ for the leptons as
well as the harder $\slashed{E_T}$-distribution of the signal compared to the background.  
These characteristics percolate through all three variables, and, albeit in a correlated fashion,
constitute important inputs in a neural network analysis, as will be reported later in this paper.

We next consider the isolation $\Delta R (= \sqrt{\Delta \eta^2 + \Delta \phi^2}$) between the two leptons. From Figure~\ref{ht_drll_ll} (right) it can be seen the peaks for signal processes are at a lower value than that of the backgrounds. The signal dileptons come from the $H^{\pm\pm}$ and thus have a higher probability of being in the same hemisphere, than in the case of the dominant background
channels. However, the $H^{\pm\pm}$ produced in a Drell-Yan process is devoid of large boost,
thus preventing the aforesaid isolation from being a very good discriminator. It nonetheless has
a role in the neural network analysis.

It is relevant to mention here
that the above kinematic distributions for BP 2 and 3 look quite similar. The reason behind this is, in both the cases the mass of the heavy Higgs states are same. On the one hand, the
lepton hardness level is controlled by the $m_{H^{\pm\pm}}$. On the other side,
$\slashed{E_T}$, too, is decided by $m_H$, though the invisible decay
of the latter takes place in different final states for the two benchmark points; 
for BP 2 it is $H \rightarrow \nu \nu$, and $H \rightarrow \chi \chi$ for BP 3.

\subsubsection{Results}

Based on the preceding observations, we have applied the following cuts on the observables. 
The events selected will have at least two jets and two same-sign dileptons($e/\mu$). The leptonic decay of $\tau$ has not been considered since its contribution is rather small.

\begin{itemize}
\item Cut 1: The invariant mass of the same-sign dileptons $m_{ll} > 400$ GeV.
\item Cut 2: Cluster transverse mass $M_{cluster} > 700 $GeV.
\item Cut 3. Scalar $p_T$ sum $H_T > $ 700 GeV.
\item Cut 4: Transverse mass $ M_T > $ 550 GeV.
\item Cut 5: $\slashed{E_T} > 300$ GeV.
\item Cut 6: $p_T$ of the leading lepton $> 250$ GeV and $p_T$ of the sub-leading lepton $> 200$ GeV.
\end{itemize}

\begin{table}[!hptb]
\begin{center}
\begin{footnotesize}
\begin{tabular}{| c | c | c | c | c | c | c | c |}
\hline
 & BP 1 & BP 2 & BP 3 & $t \bar t$ & $W$ + jets & $t \bar t W$ & WZ  \\
\hline
$\sigma(fb)$ & 0.12 & 0.19 & 0.11 & $3.09 \times 10^5$ & $2.8 \times 10^7$ & 9.77  &  355.10  \\
\hline
Cut 1 & 99.3\% & 99.6\% & 99.6\% & 0.2\% & 0.15\% & 2.1\% & 1.8\% \\
\hline
Cut 2 & 99.2\% & 99.6\% & 99.6\% & 0.1\% & 0.08\% & 1.7\% & 1.3\% \\
\hline
Cut 3 & 96.8\% & 99.2\% & 99.1\% & 0.06\% & 0.03\% & 0.9\% & 0.4\% \\
\hline
Cut 4 & 96.8\% & 99.2\% & 99.1\% & 0.05\% & 0.026\% & 0.8\% & 0.3\% \\
\hline
Cut 5 & 73.5\% & 87.4\% & 87.6\% & $0.01$\% & 0.003\% & 0.07\% & 0.04\% \\
\hline
Cut 6 & 40.2\% & 62.5\% & 62.4\% & $0.002$\% & 0.0009\% & 0.01\% & 0.005\% \\
\hline
\end{tabular}
\end{footnotesize}
\caption{Signal and background efficiencies after applying various cuts for case I at 14 TeV. The cross-sections are calculated at NLO.}
\label{llcuts}
\end{center}
\end{table}

Table~\ref{llcuts} shows the cut-flow for the signal and the background for case I, yielding
a fair indication of the efficiency of each cut. In Table~\ref{significance_ll} we calculate the projected significance (${\cal S})$ for each benchmark point for the 14 TeV LHC with 3000 $fb^{-1}$. The significance ${\cal S}$ is defined as 

\begin{equation}
{\cal S} = \sqrt{2 [(S+B) \text{Log}(1+\frac{S}{B}) - S]}
\label{significance}
\end{equation}

Where $S$ and $B$ are the number of signal and background events surviving the 
succession of cuts.

\begin{table}[!hptb]
\begin{center}
\begin{footnotesize}
\begin{tabular}{| c | c |}
\hline
BP & $
{\cal S}$   \\
\hline
BP 1  &  3.4 $\sigma$  \\
\hline
BP 2  & 8.3 $\sigma$  \\
\hline
BP 3  & 5.0 $\sigma$  \\
\hline
\end{tabular}
\end{footnotesize}
\caption{Signal significance for the benchmark points at 14 TeV with ${\cal L}$ = 3000 $fb^{-1}$ for case I. }
\label{significance_ll}
\end{center}
\end{table}

We can see from Table~\ref{significance_ll} that for BP 2 the largest significance is predicted. 
Although in BP 2 the production cross-section for $H^{\pm\pm}H^{\mp}$ is smaller compared to 
that in BP 1, BP 2 has large invisible branching ratio(mostly $H \rightarrow \nu \nu$) as well as large Br($H^{\pm\pm} \rightarrow \ell^{\pm}\ell^{\pm}$) since it corresponds to the  smallest triplet VEV among the
three benchmarks. On the other hand, BP 1 has smaller Br($H^{\pm\pm} \rightarrow \ell^{\pm}\ell^{\pm}$) because of larger triplet VEV, and consequently smaller $\Delta L =2$ interaction strengths
(in order to conform to the neutrino mass limits). Therefore, even with large invisible branching fraction for $H \rightarrow \chi \chi$ this BP suffers from lower overall rate. In case of BP 3, Br($H \rightarrow \chi \chi$) and Br($H \rightarrow \nu \nu$) are comparable, the smaller Br($H^{\pm\pm} \rightarrow \ell^{\pm}\ell^{\pm}$) due to smaller triplet VEV makes this BP a  little more challenging than BP 2
from the experimental point of view. 
Moreover, the masses of the heavy states $H^{\pm\pm}, H^{\pm}$ and $H$ are larger 
in BP 2 and 3, as 
compared to BP 1. Thus one has better handle on the signal separation process,
using the variables discussed already.


\subsection{Case II}\label{sub5.2}

For relatively large ($\gsim 10^{-3}$ GeV) triplet VEV, the $H^{\pm\pm}$ produced in the Drell-Yan process will decay into a pair of same-sign $W$ bosons. The leptonic decay of the produced $W$-bosons once more gives rise to same-sign dileptons along with $\slashed{E_T}$, but
without any dilepton invariant mass peak. It is profitable
to latch on to hadronic decays of the $W$ coming from the associated $H^{\pm}$ decaying into $HW^{\pm}$ final state.
 When the above decay is kinematically suppressed, the $H^{\pm}$ will decay into $W^{\pm}h$ or $W^{\pm}Z$ final states, empowered by the relatively higher triplet VEV. The subsequent invisible decay of $H$ will be a tell-tale signature of dark matter, the $\nu\nu$ mode being
 suppressed by the Yukawa coupling in this case.  

The sources of backgrounds here are the same as in case I. However, the fact that same-sign dileptons in this case do not come from a single source causes somewhat different kinematical features compared to case I, as we will see below.

\subsubsection{Distributions}

\begin{figure}[!hptb]

\includegraphics[width=8.8cm, height=7cm]{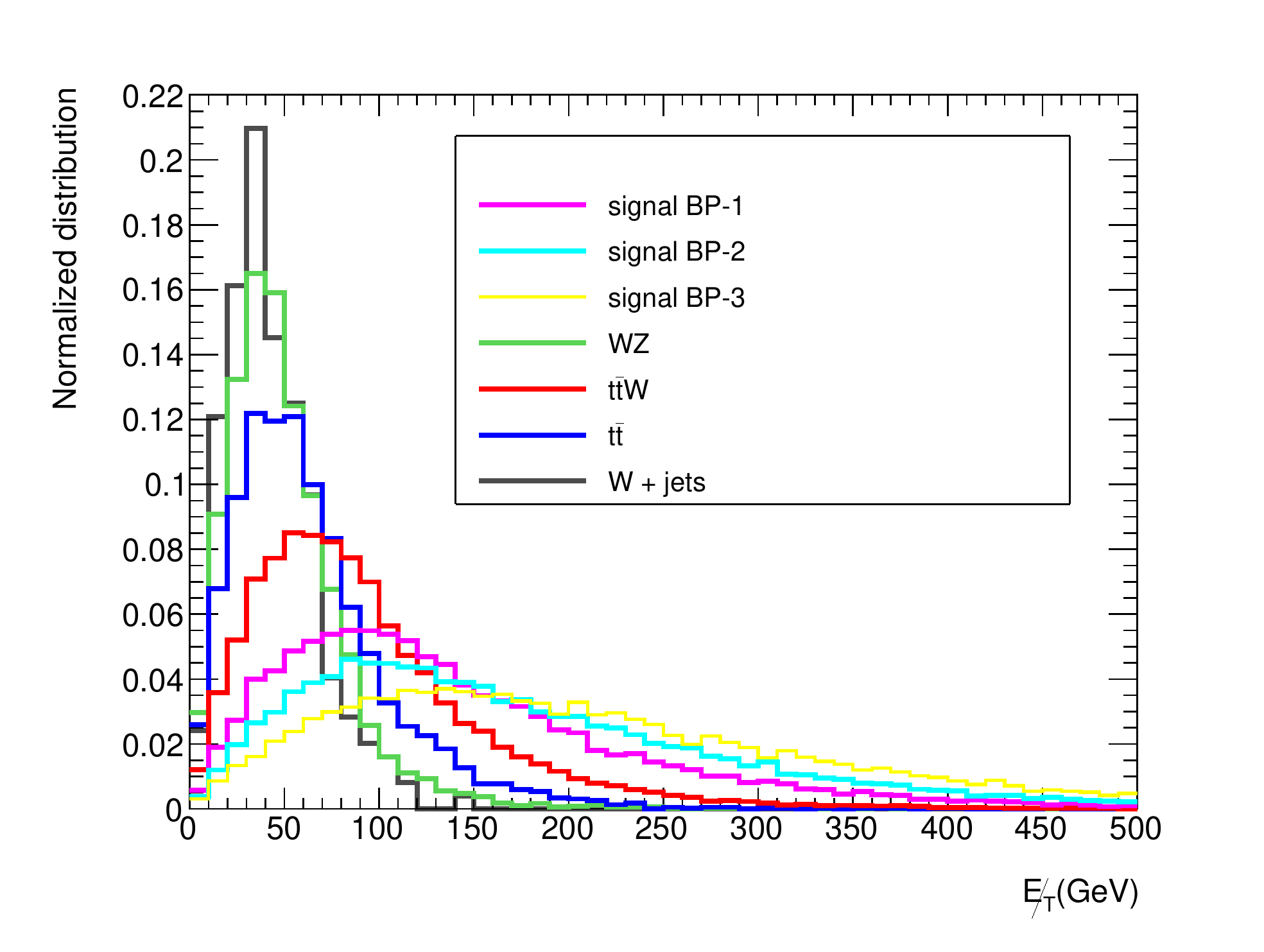}
\includegraphics[width=8.8cm, height=7cm]{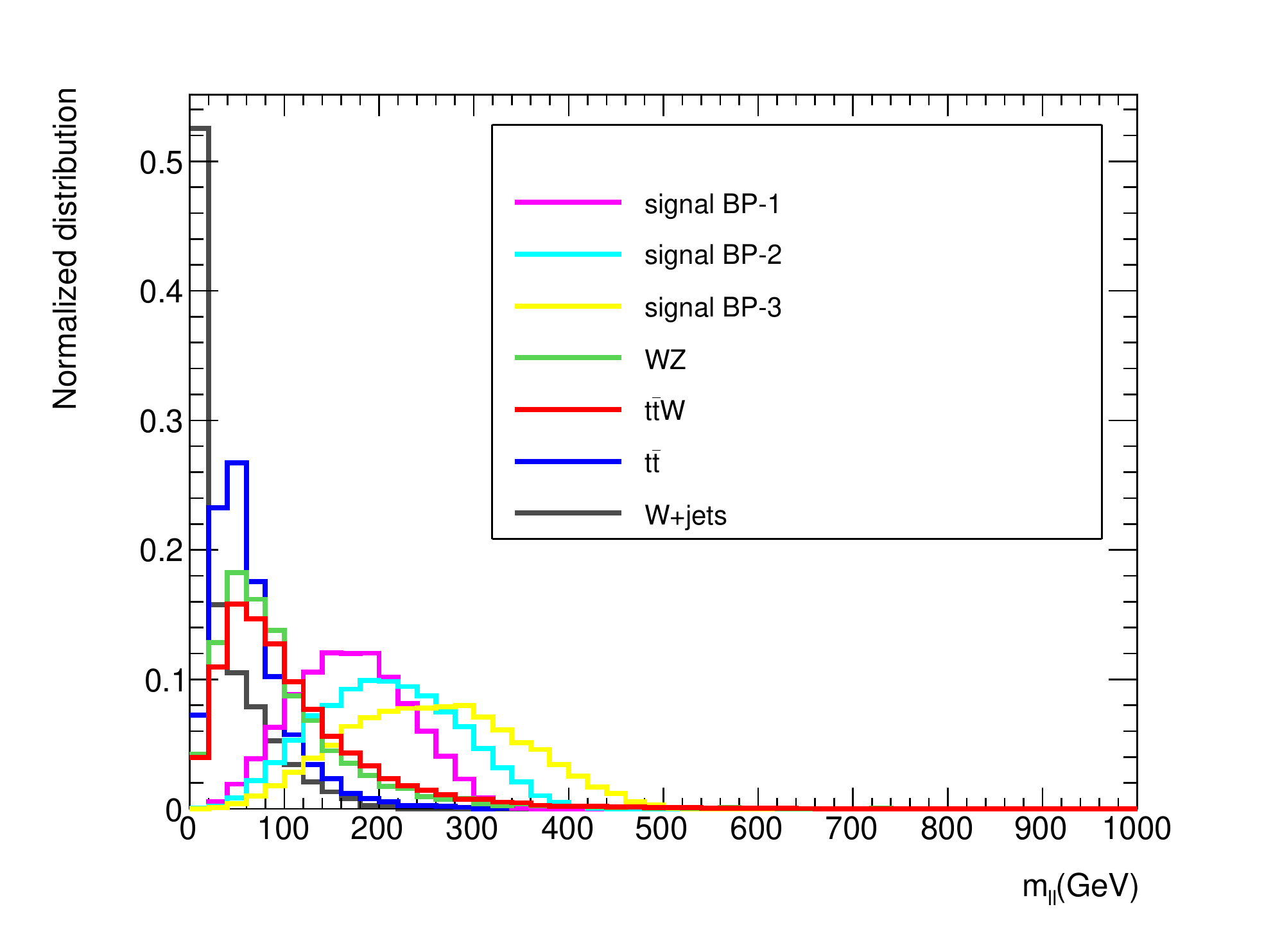}

\caption{Distribution of $\slashed{E_T}$(left) and invariant mass(right) of same-sign dileptons for the three signal BPs and backgrounds in case II.}
\label{misspt_invll_ww}
\end{figure}

\begin{figure}[!hptb]

\includegraphics[width=8.8cm, height=7cm]{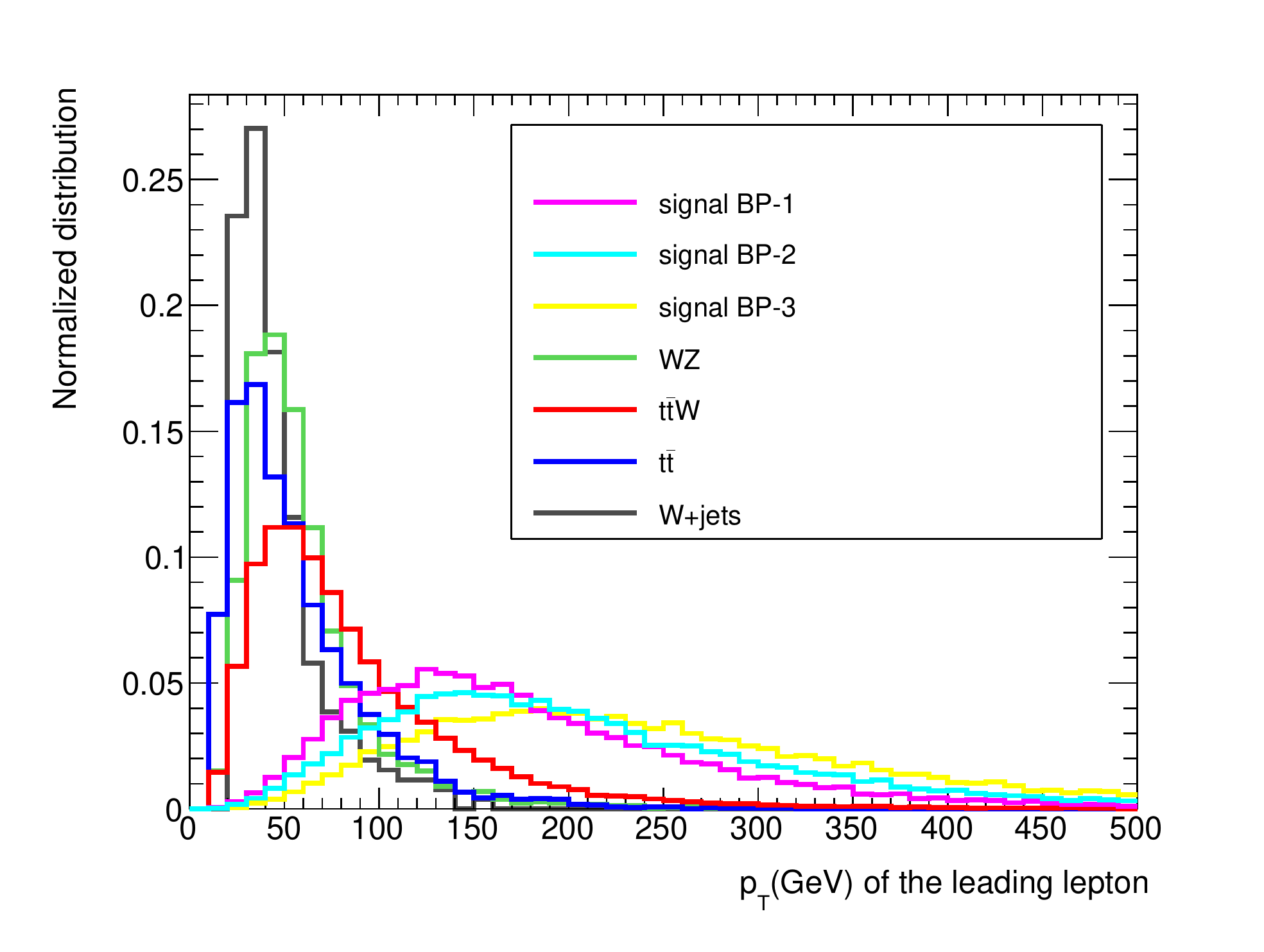}
\includegraphics[width=8.8cm, height=7cm]{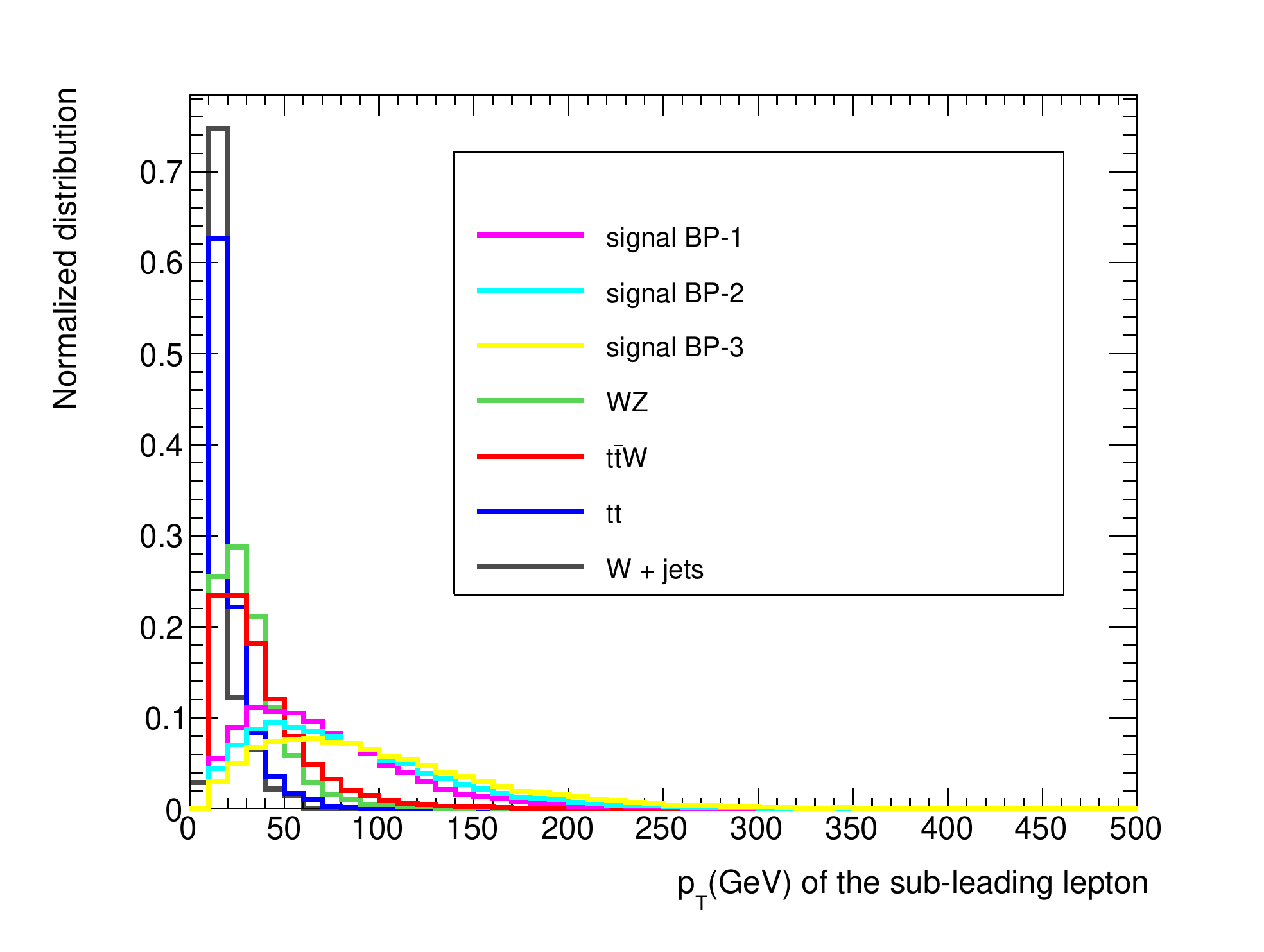}

\caption{Distribution of transverse momenta of the leading(left) and sub-leading(right) leptons for the three signal BPs and backgrounds in case II.}
\label{ptl1_ptl2_ww}
\end{figure}

\begin{figure}[!hptb]

\includegraphics[width=8.8cm, height=7cm]{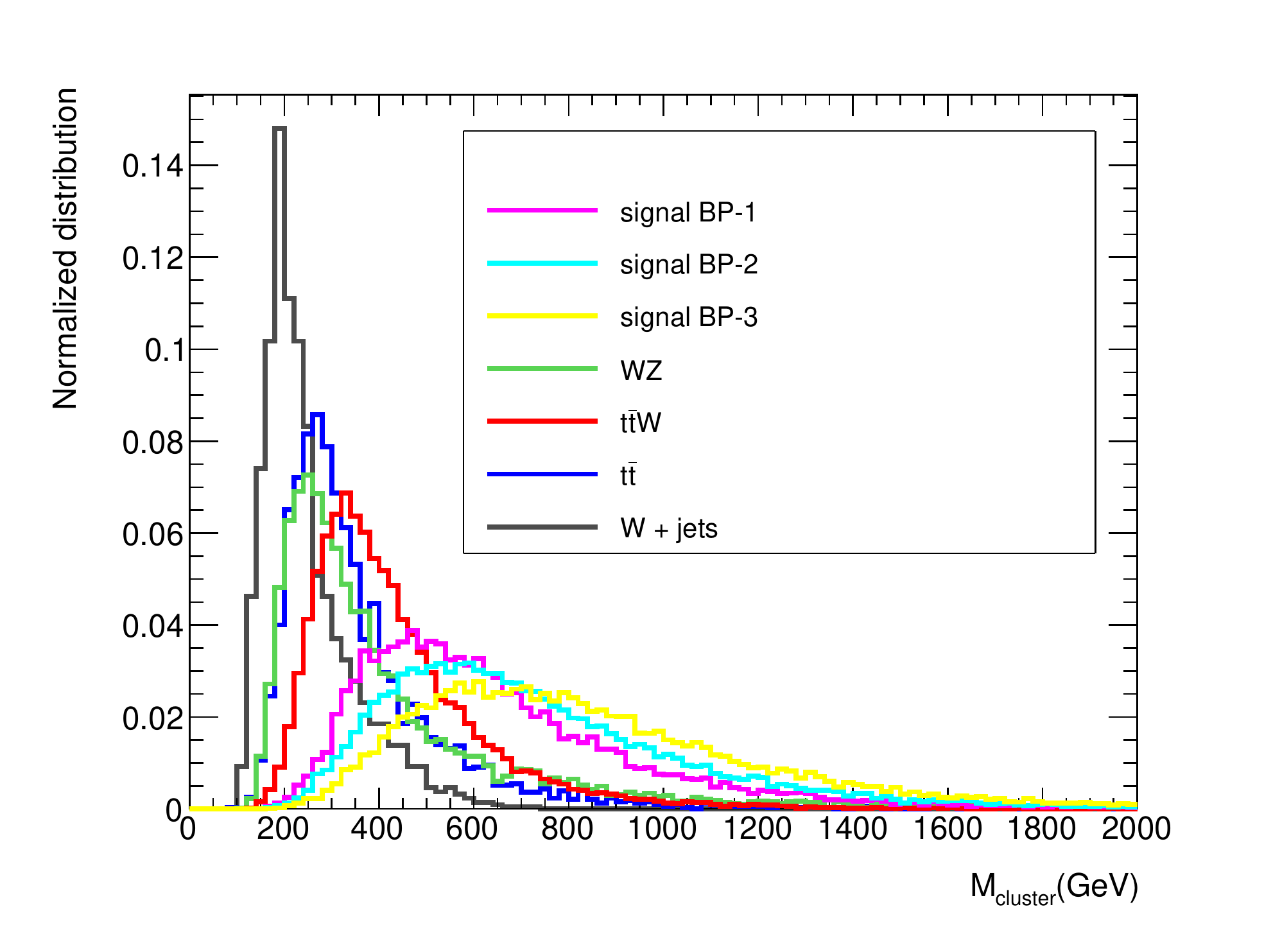}
\includegraphics[width=8.8cm, height=7cm]{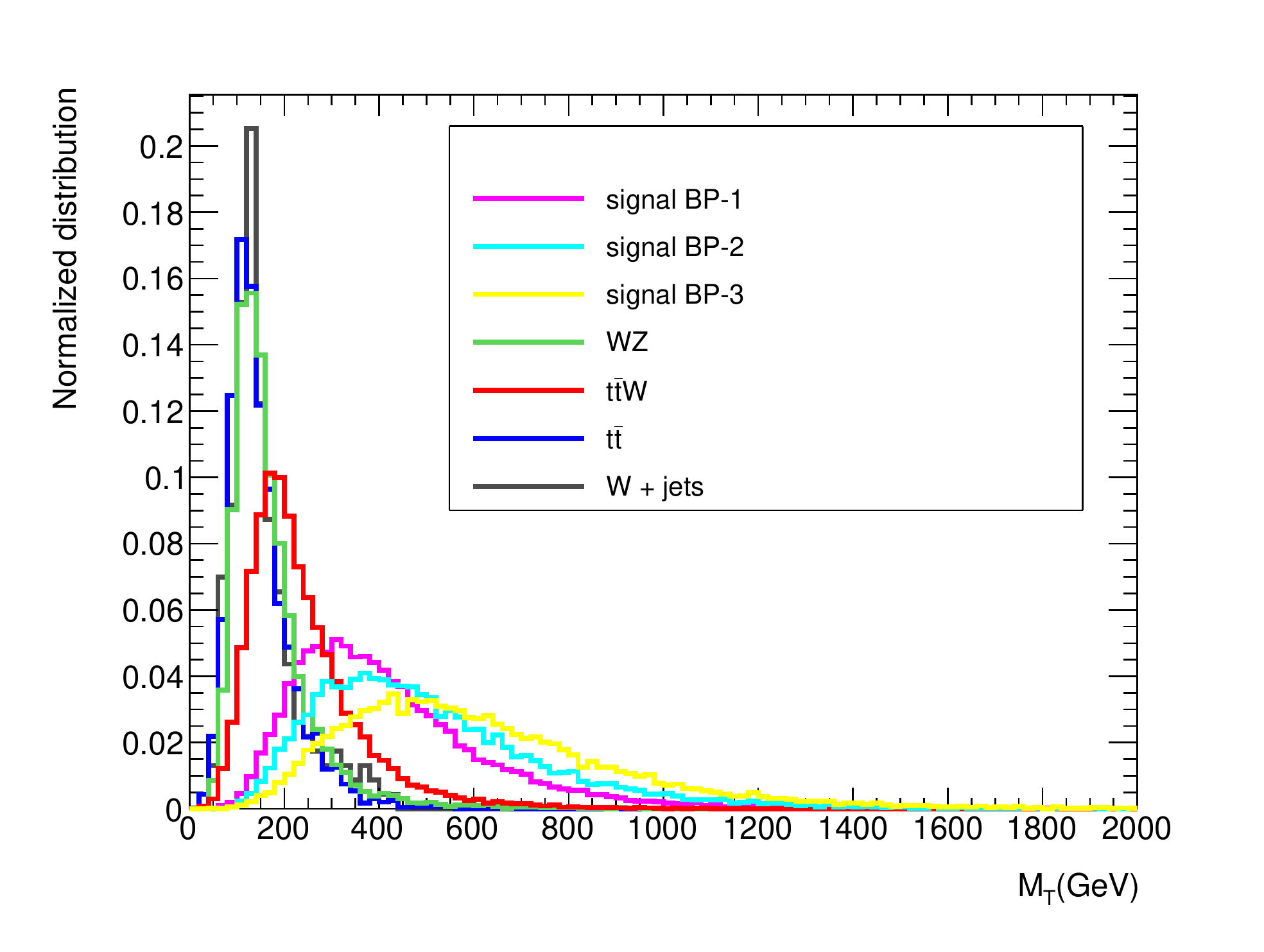}

\caption{Distribution of cluster transverse mass(left) and transverse mass(right) for the three signal BPs and backgrounds in case II.}
\label{mcl_mtr_ww}
\end{figure}

\begin{figure}[!hptb]

\includegraphics[width=8.8cm, height=7cm]{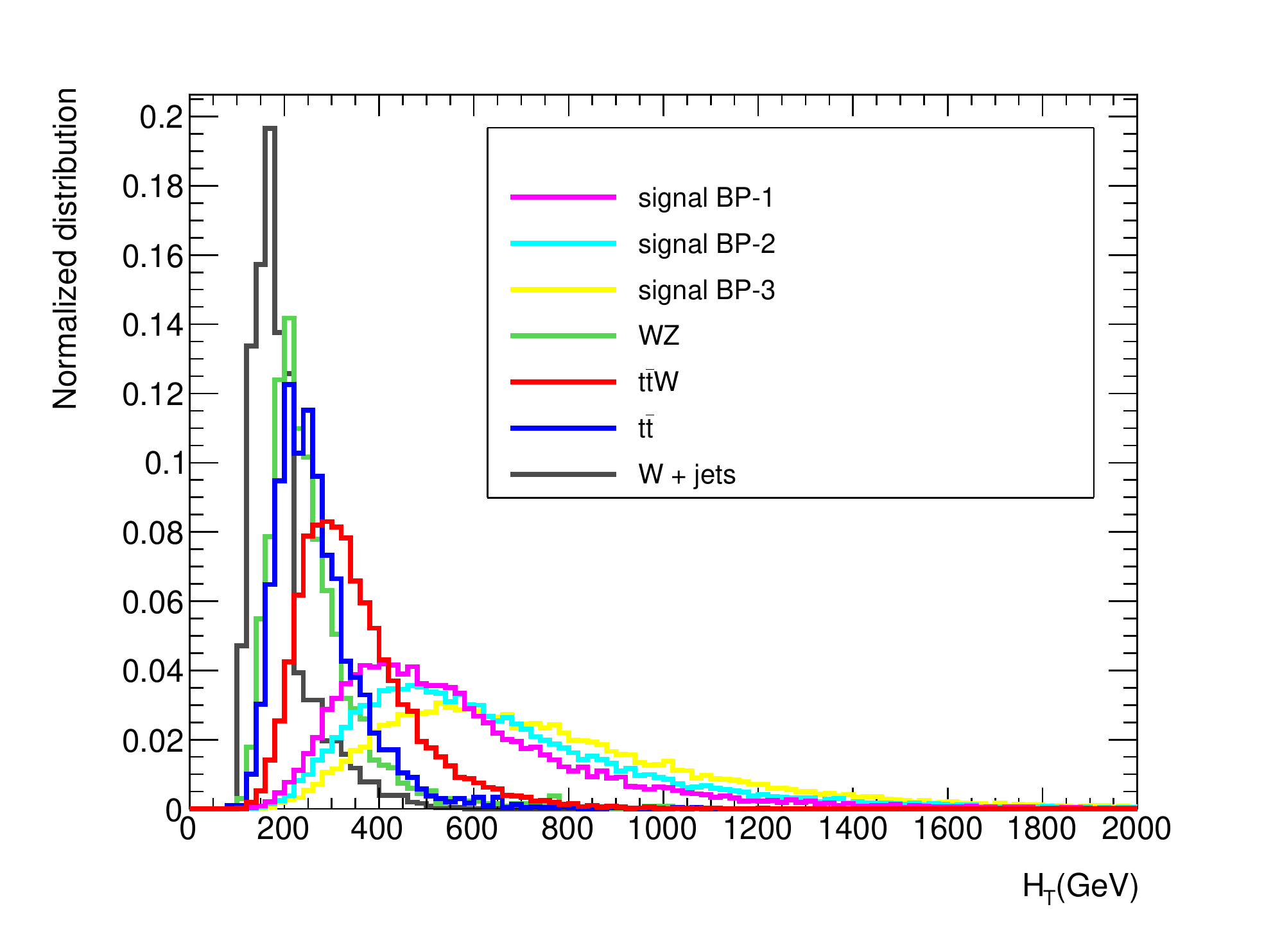}
\includegraphics[width=8.8cm, height=7cm]{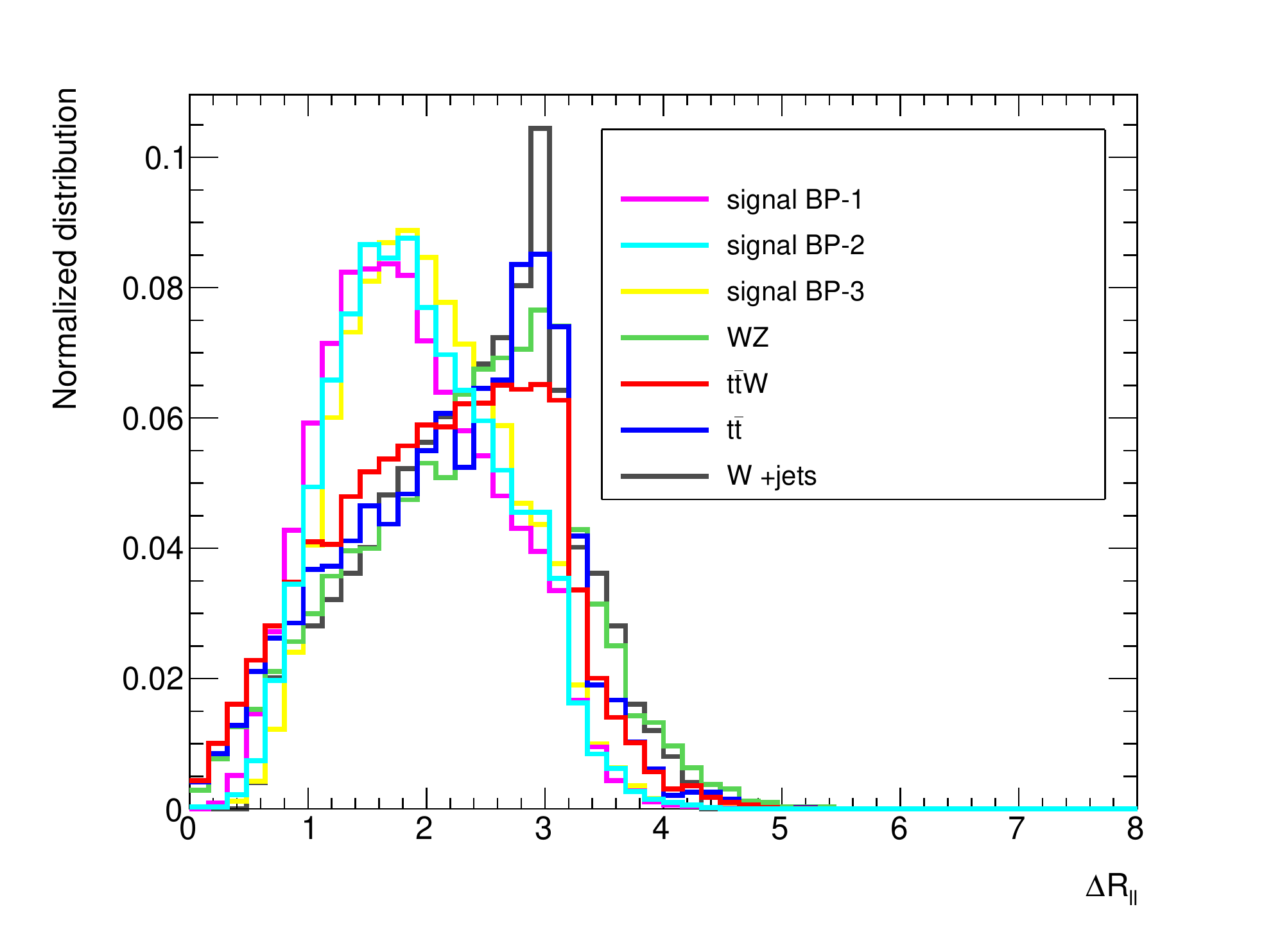}

\caption{Distribution of $H_T$ and $\Delta R$ between leading and sub-leading leptons for the three signal BPs and backgrounds in case II.}
\label{ht_drll_ww}
\end{figure}

In Figure~\ref{misspt_invll_ww} (left) we plot the $\slashed{E_T}$ distribution in the final state. We can see that for the signal processes, the distribution peaks at lower values than case I even when masses of heavy Higgses are in similar regions. This is because the source of neutrinos here are
the two boosted same-sign $W$-bosons, which occur in the hemisphere opposite to the one where
the $H$ emanates, thus enabling the cancellation of missing transverse momenta.

 Figure~\ref{misspt_invll_ww} (right) shows the invariant mass distribution of the same-sign dilepton pair. The peak in this distribution also shifts to a lower value compared to case I, largely because
 of the reduced individual energy share of each participating lepton. 
 The signal $p_T$ distributions, too, peak at a lower values compared to case I,
 as seen in Figure~\ref{misspt_invll_ww}. A similar fate also awaits
 $M_{cluster}, M_T$ and $H_T$, as seen from Figures~\ref{mcl_mtr_ww} and~\ref{ht_drll_ww}. 
Along with similar, and less consequential isolations as in  Figure~\ref{ht_drll_ww} (right),
these features make the statistical significance relatively modest in Case II.

\subsubsection{Results}

Gaining some insight into the kinematics of the final state particles in signal and background processes, we apply various cuts on the relevant observables and perform a cut-based analysis. The events with exactly two same-sign dileptons and at least two jets are selected.
The following cuts have been applied in succession on both signal and background events.

\begin{itemize}
\item Cut 1: The invariant mass of the same-sign dileptons $m_{ll} > 150$ GeV.
\item Cut 2: Cluster transverse mass $M_{cluster} > 500 $GeV.
\item Cut 3. Scalar $p_T$ sum $H_T > $ 500 GeV.
\item Cut 4: Transverse mass $ M_T > $ 500 GeV.
\item Cut 5: $\slashed{E_T} > 250$ GeV.
\item Cut 6: $p_T$ of the leading lepton $> 200$ GeV and $p_T$ of the sub-leading lepton $> 100$ GeV. 
\end{itemize}

\begin{table}[!hptb]
\begin{center}
\begin{footnotesize}
\begin{tabular}{| c | c | c | c | c | c | c | c |}
\hline
 & BP 1 & BP 2 & BP 3 & $t \bar t$ & $W$ + jets & $t \bar t W$ & $WZ$ \\
\hline
$\sigma(fb)$ & 0.79 & 0.18 & 0.10 & $3.09 \times 10^5$ & $2.8 \times 10^7$ & 9.77  &  355.10  \\
\hline
Cut 1 & 62.0\% & 77.0\% & 88.0\% & 4.8\% & 3.1\% & 28.2\% & 24.4\% \\
\hline
Cut 2 & 47.0\% & 64.0\% & 78.2\% & 3.0\% & 1.8\% & 18.2\% & 11.2\% \\
\hline
Cut 3 & 39.0\% & 55.0\% & 70.0\% & 0.7\% & 0.42\% & 12.0\% & 4.0\% \\
\hline
Cut 4 & 23.0\% & 39.0\% & 57.3\% & 0.2\% & 0.1\% & 4.0\% & 1.4\% \\
\hline
Cut 5 & 8.8\% & 20.0\% & 31.8\% & 0.04\% & 0.014\% & 0.8\% & 0.3\% \\
\hline
Cut 6 & 4.7\% & 10.0\% & 17.2\% & 0.01\% & 0.004\% & 0.1\% & 0.05\% \\
\hline
\end{tabular}
\end{footnotesize}
\caption{Signal and background efficiencies after applying various cuts for case II at 14 TeV. The cross-sections are calculated at NLO.}
\label{wwcuts}
\end{center}
\end{table}

In Table~\ref{wwcuts} we present the cut-flow for signal and backgrounds for case II. Finally, Table~\ref{significance_ww} contains the projected signal significance for the three benchmarks
for 14 TeV LHC with 3000 $fb^{-1}$ data. The significance ${\cal S}$ is defined in Equation~\ref{significance}.

\begin{table}[!hptb]
\begin{center}
\begin{footnotesize}
\begin{tabular}{| c | c |}
\hline
BP & $
{\cal S}$   \\
\hline
BP 1  &  2.0 $\sigma$  \\
\hline
BP 2  & 1.0 $\sigma$  \\
\hline
BP 3  & 1.1 $\sigma$  \\
\hline
\end{tabular}
\end{footnotesize}
\caption{Signal significance for the benchmark points at 14 TeV with ${\cal L}$ = 3000 $fb^{-1}$ for case II.}
\label{significance_ww}
\end{center}
\end{table}

We can see from Table~\ref{significance_ww} that only BP 1 will have substantial significance at 3000 $fb^{-1}$ luminosity. The major reason behind that is large production cross-section helped by 
comparatively low heavy Higgs masses. Moreover, this benchmark also has 
all relevant  branching fractions, namely, those for $H \rightarrow \chi\chi, 
H^{\pm} \rightarrow HW^{\pm}$  and $H^{\pm \pm} \rightarrow W^{\pm} W^{\pm}$, working in favour
of the signal. It has Br($H \rightarrow \chi \chi \approx 90\%$). On the other hand, BP 1 has the lowest triplet VEV among the three BPs. In this case $H^{\pm}$ decays mostly to $HW^{\pm}$ final state. For BP 2, however,  other decay channels like $hW^{\pm}, t \bar b$ etc open up, hence the Br$(H^{\pm} \rightarrow HW^{\pm}$) falls (27\% in case of BP 2 as this channel has the largest VEV). Therefore, although BP 2 and 3 have better separation between signal and background owing to large heavy Higgs masses, the low cross-sections and branching fractions make such regions in the
parameter space somewhat challenging. Keeping this in mind, the remaining part of our
investigation goes beyond rectangular cuts.

\subsection{W-boson fusion}

As an alternative channel, one may think of $W$-boson fusion, since it provides the useful forward jets tag. Here a relevant production channel could be $p p \rightarrow H^{\pm\pm} H$ + two forward jets along with $H$ decaying into the invisible channel, and leading to same-sign dilepton + $\slashed{E_T}$ in the rapidity interval between the forward jets. On actual calculation, however, it is found that even the most optimistic benchmarks lead to production cross-section $\approx 10^{-2} fb$. The event rate after factorizing in the decay branching ratios and applying various selection criteria thus becomes rather small even for the HL-LHC. We therefore do not enter into detailed analysis of this channel.

\section{Results with gradient boosting and neural networks}\label{sec6}

Having performed the rectangular cut-based analysis for same-sign dilepton + $\slashed{E_T}$ signal, we see that some benchmark points yield very good signal significance at the HL-LHC. Therefore they will be easily detectable at the future run. However, there are some benchmarks which predict rather poor signal significance in a cut-based analysis. Specifically, BP 2 and 3 of the scenario with $H^{\pm\pm} \rightarrow W^{\pm} W^{\pm}$ yield very low significance, as seen in Table~\ref{significance_ww}. The main reason behind this is the comparatively low production cross-section and branching ratio in this case. Moreover, the absence of a same-sign dilepton peak makes it somewhat challenging in case II. Taking this issue into consideration we move towards a more sophisticated analysis using packages based on {\bf Gradient boosting (XGBoost)~\cite{Chen:2016btl} and Artificial neural network (ANN)~\cite{Teodorescu:2008zzb}} techniques. Their usefulness has been widely demonstrated~\cite{Baldi:2014kfa,Woodruff:2017geg,Oyulmaz:2019jqr,Bhattacherjee:2019fpt} including studies in the Higgs sector~\cite{Hultqvist:1995ibm,Field:1996rw,Bakhet:2015uca,Dey:2019lyr,Lasocha:2020ctd}.
In this section we will explore the possibility of improvement of our analysis using these techniques. In particular for ANN we have used   the toolkit Keras~\cite{keras}. We perform the analysis for both case I and II and also make a comparative study of the performance of ANN and XGBoost in the two cases. In Table~\ref{featurevar} we list all relevant variables these being a total of 12 such feature variables in the analysis.

\begin{table}[htpb!]
\centering
 \begin{tabular}{||c | c||} 
 \hline
 Variable & Definition \\ [0.5ex] 
 \hline\hline
 $P^{l_1}_{T}$ & Transverse momentum of the leading lepton \\ 
 $P^{l_1}_{T}$ & Transverse momentum of the sub-leading lepton \\
 $E^{miss}_{T}$ & Missing transverse energy \\
 $N_{j}$ & No of jets in the event \\
 $m_{ll}$ & Invariant mass of the same-sign dilepton pair \\
 $P^{j_1}_{T}$ & Transverse momentum of the leading jet \\ 
 $P^{j_1}_{T}$ & Transverse momentum of the sub-leading jet \\
 $m_{jj}$ & Invariant mass of the jets \\
 $m_{cluster}$ & The cluster transverse mass\\
 $m_{transverse}$ & Transverse mass \\
 $H_T$ & Scalar sum of $p_T$ of all the final state particles \\
 $\Delta R_{ll}$ & $\Delta R$ between two leptons \\ [1ex] 
 \hline
 \end{tabular}
 \caption{Feature variables for training in the XGBoost and ANN analysis.}
  \label{featurevar}
\end{table}

\begin{figure}[!hptb]

\includegraphics[width=12.8cm, height=10cm]{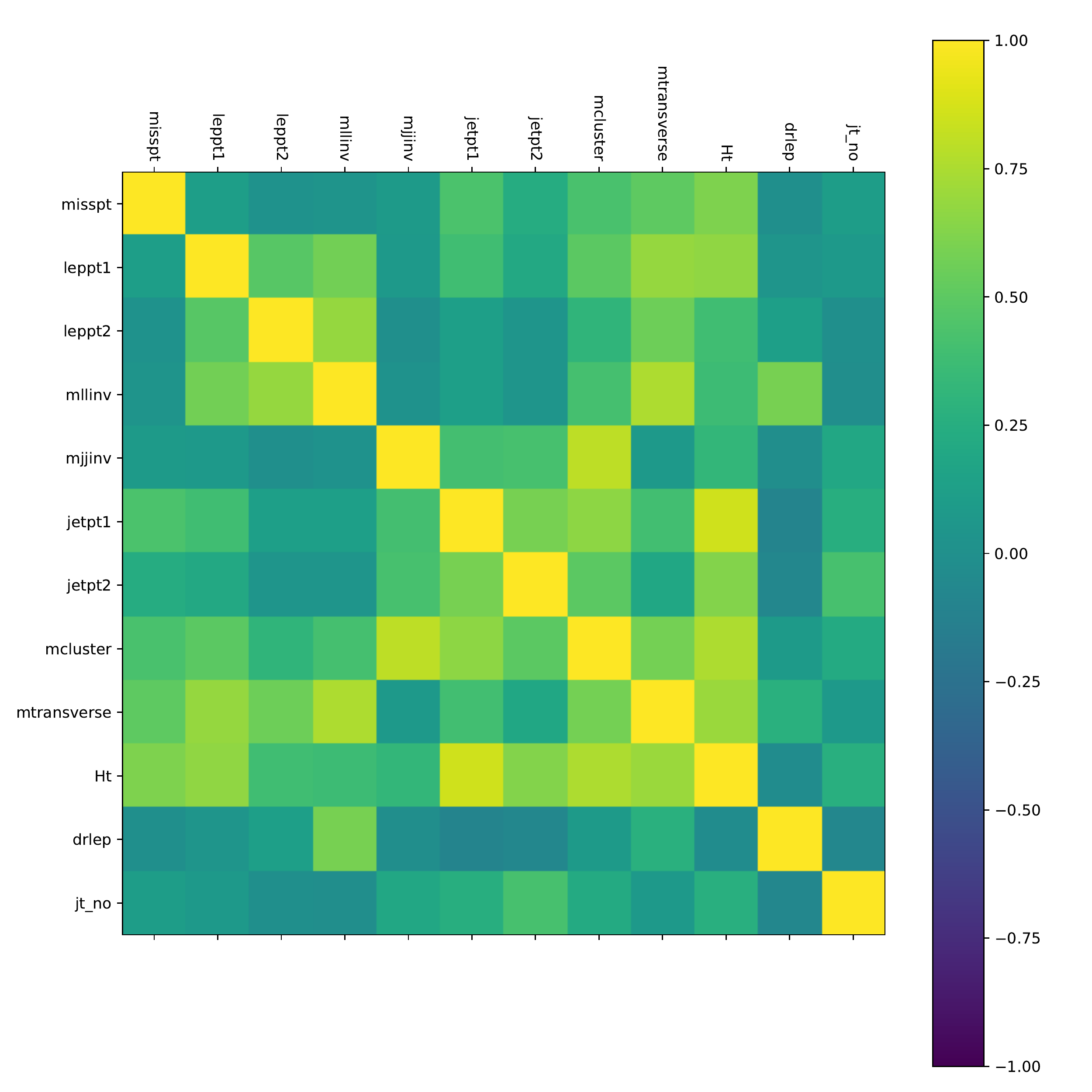}
\centering

\caption{Correlation matrix between the feature variables using XGBoost.}
\label{correlation}
\end{figure}

\begin{figure}[!hptb]

\includegraphics[width=8.8cm, height=7cm]{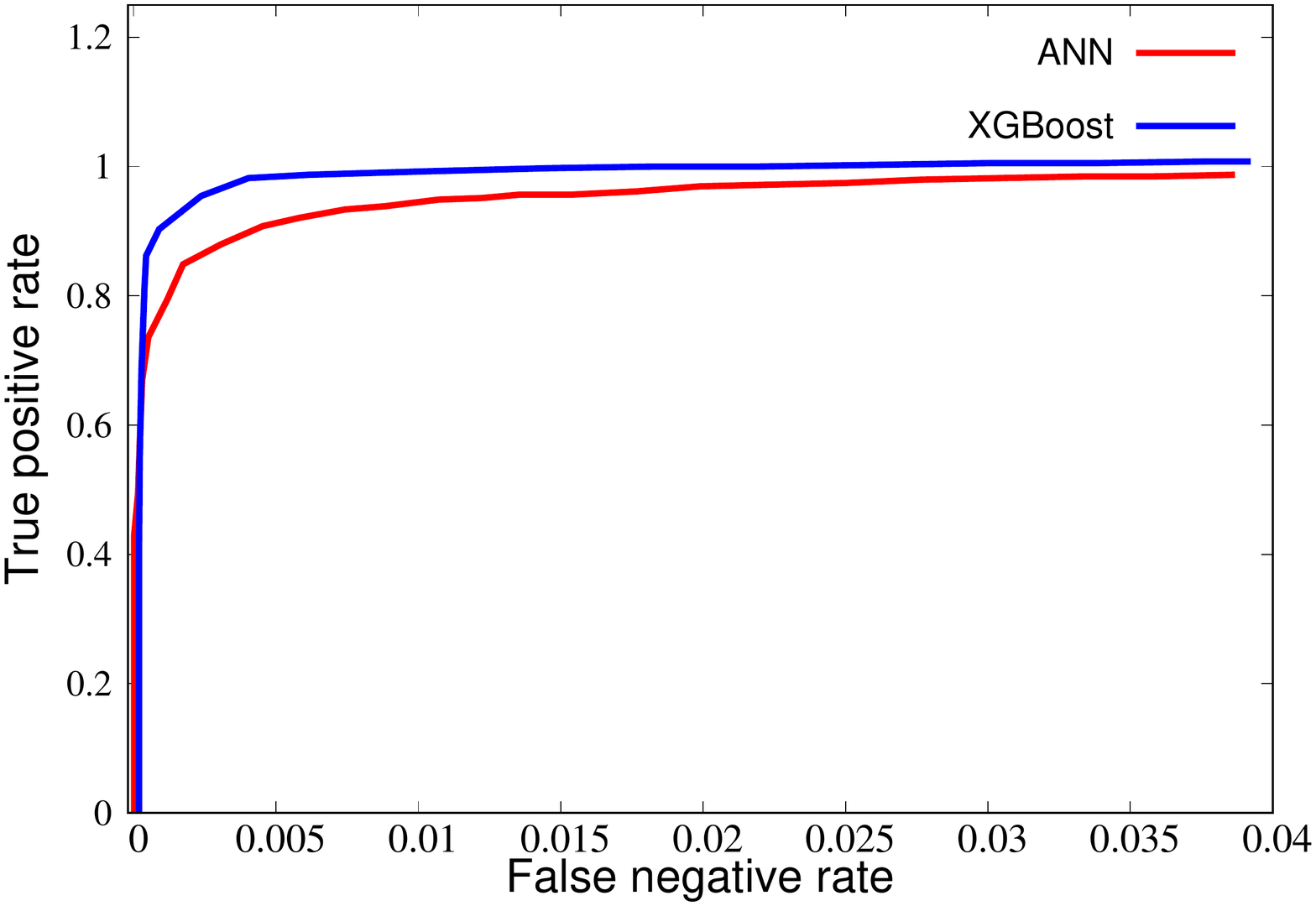}
\includegraphics[width=8.8cm, height=7cm]{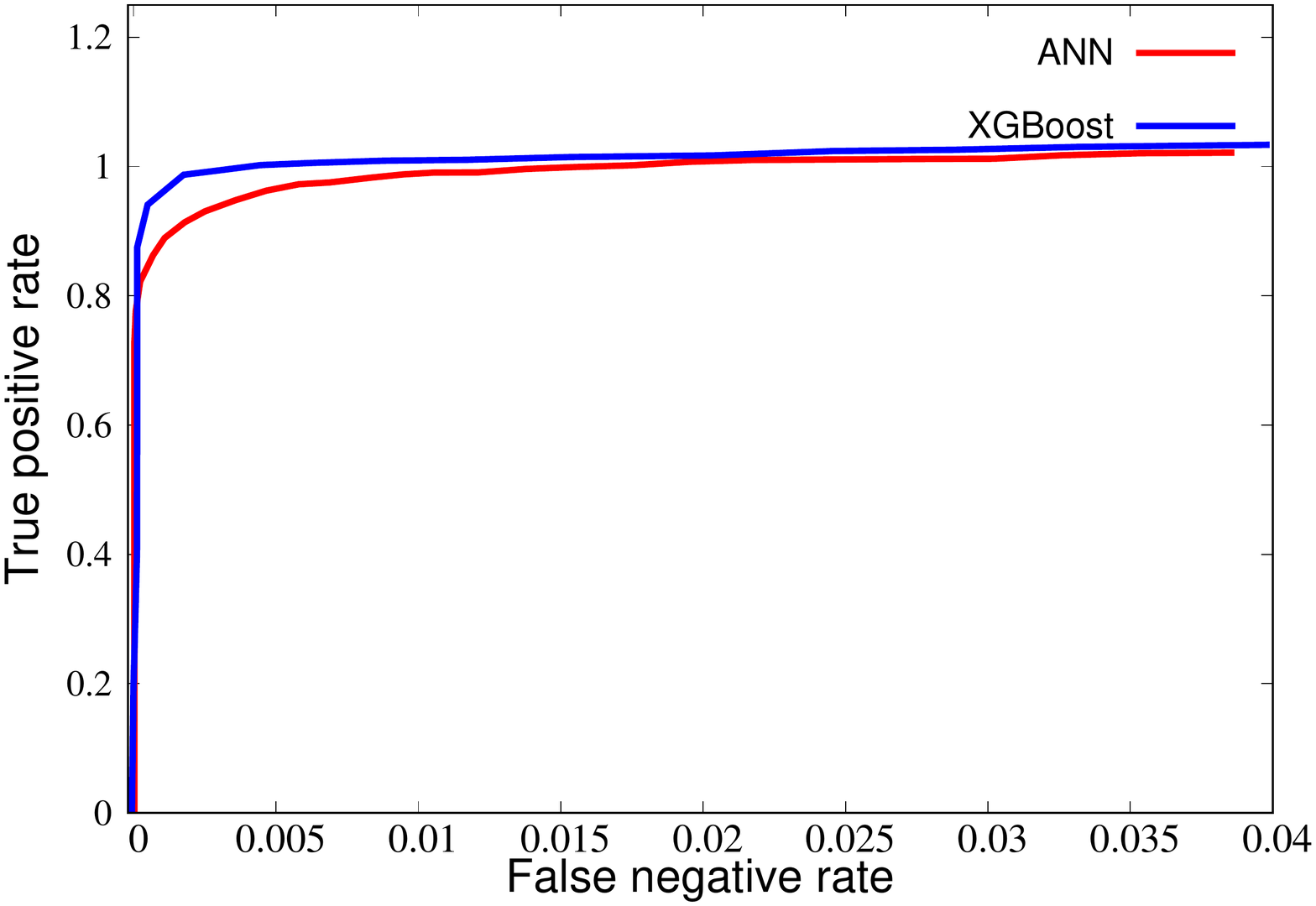} \\

\includegraphics[width=8.8cm, height=7cm]{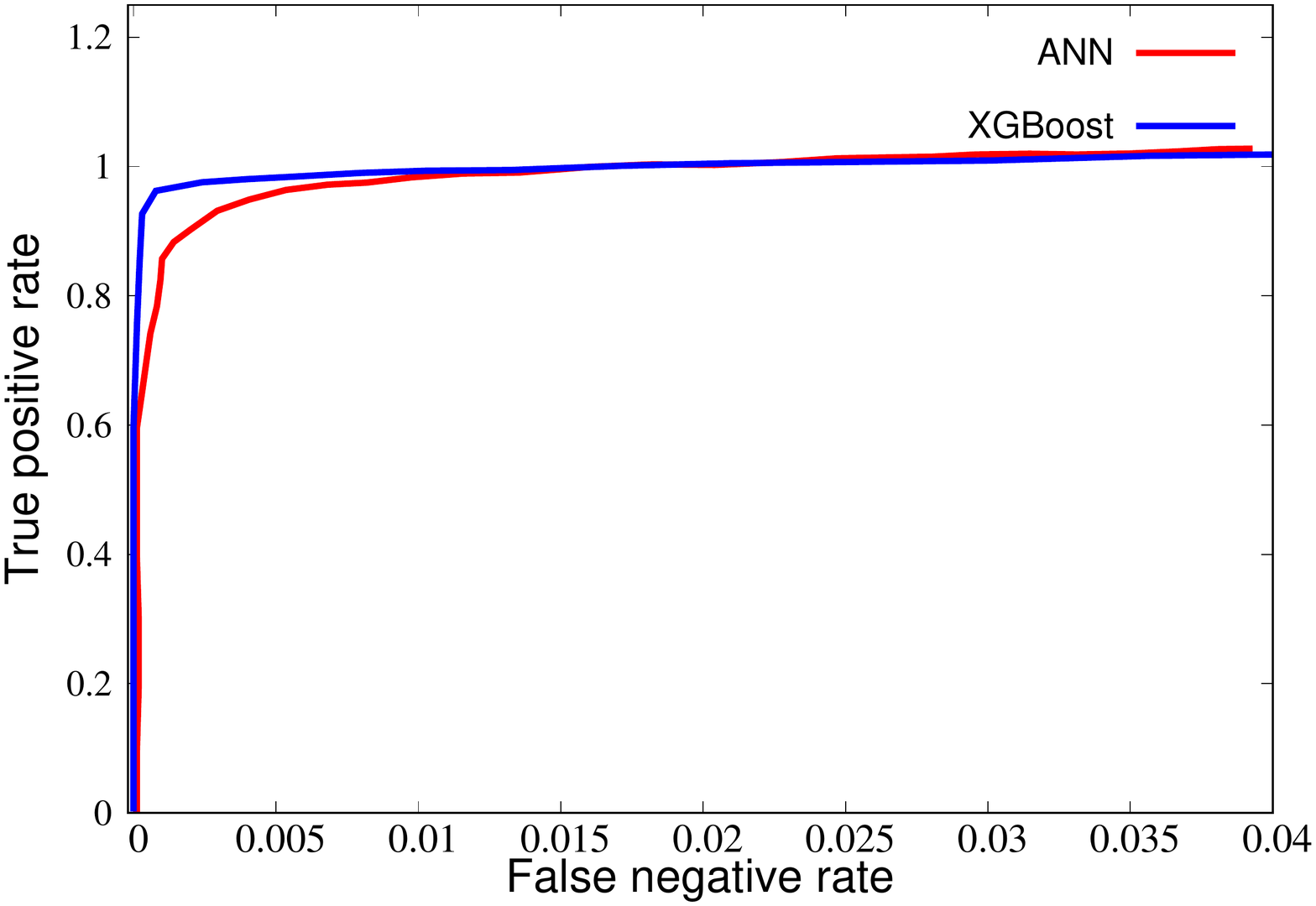}
\centering

\caption{ROC curves of BP 1 (top left), BP 2 (top right) and BP 3 (bottom centre) in case I with ANN and XGBoost.}
\label{rocll}
\end{figure}

\begin{figure}[!hptb]

\includegraphics[width=8.8cm, height=7cm]{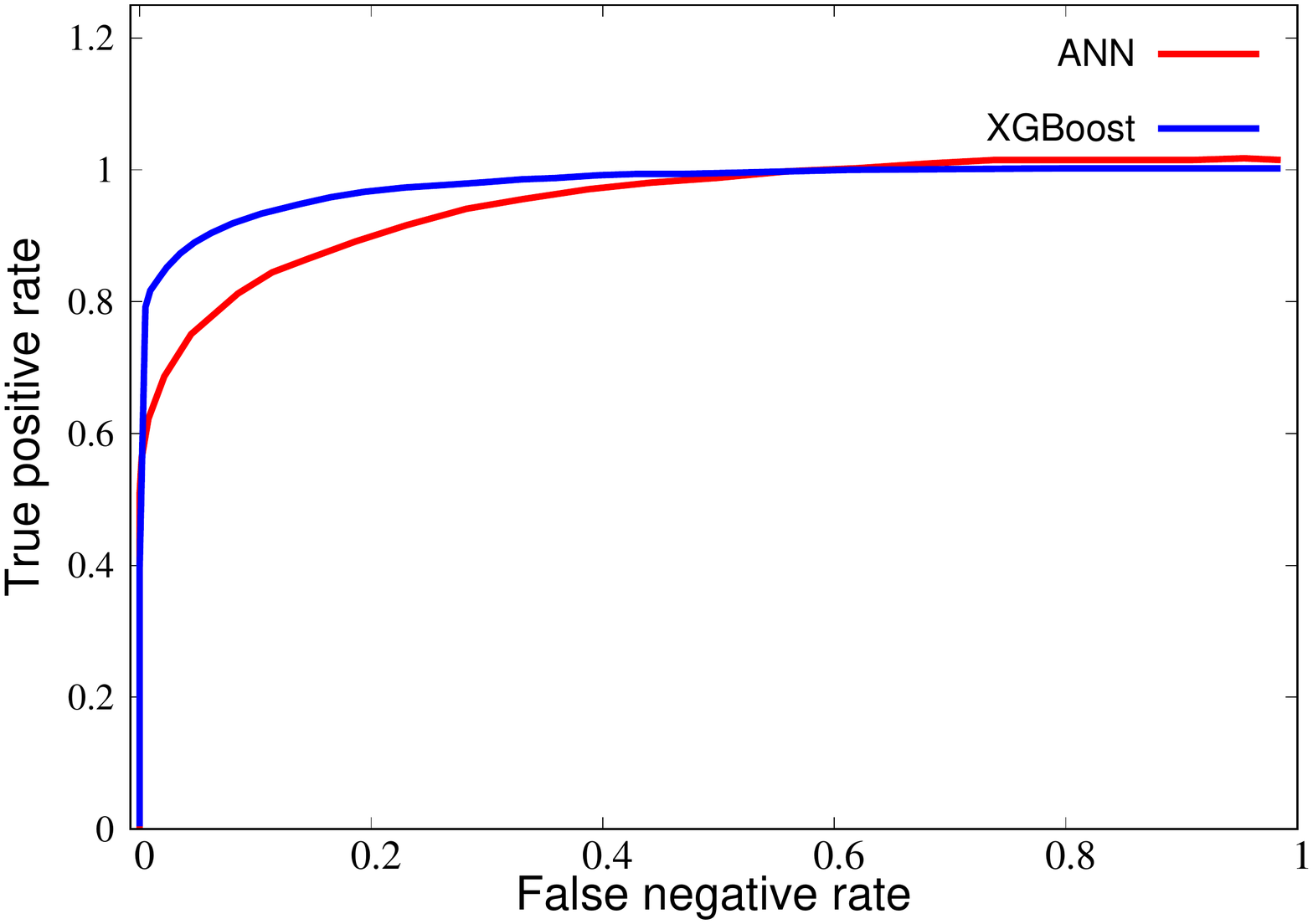}
\includegraphics[width=8.8cm, height=7cm]{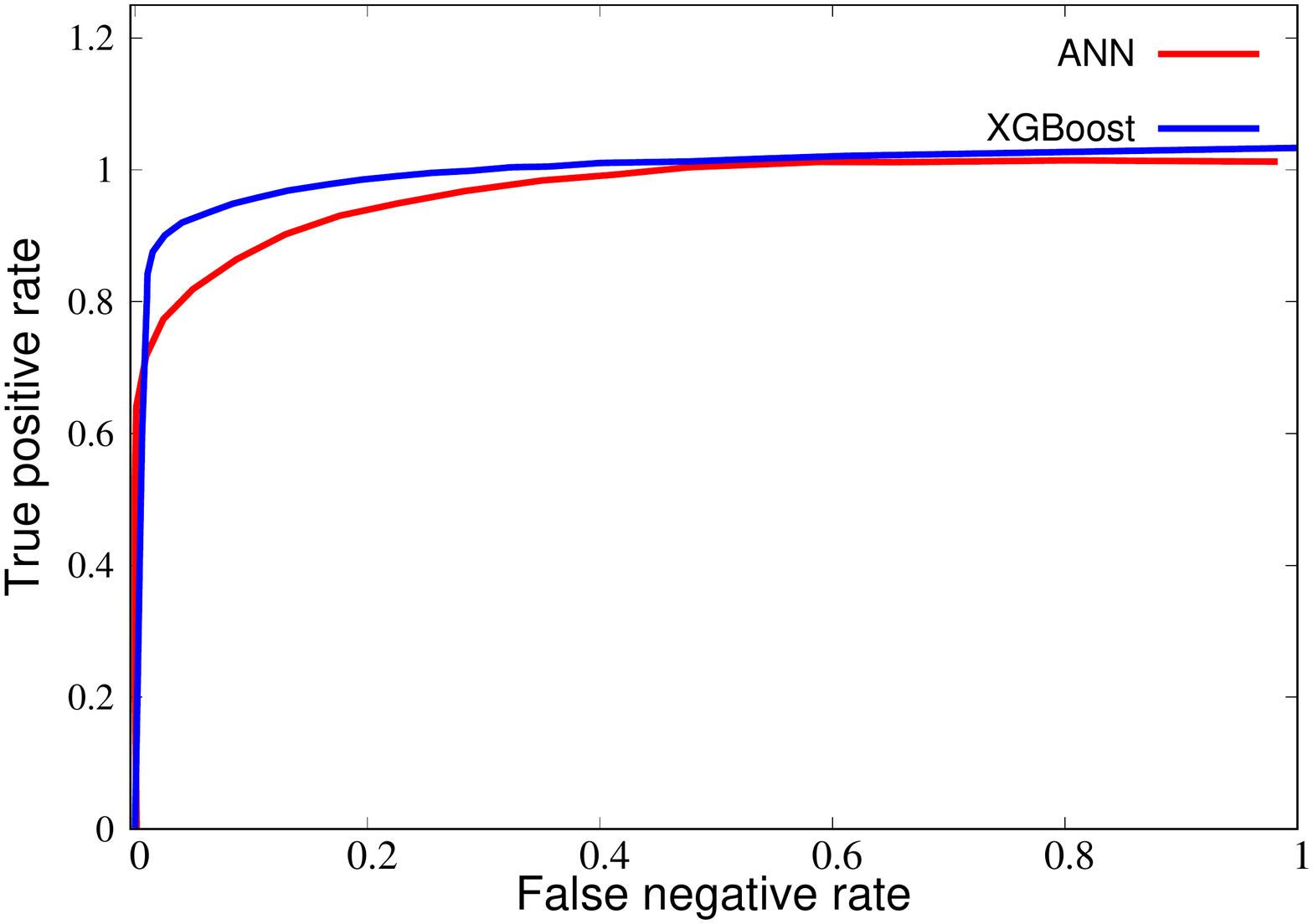} \\

\includegraphics[width=8.8cm, height=7cm]{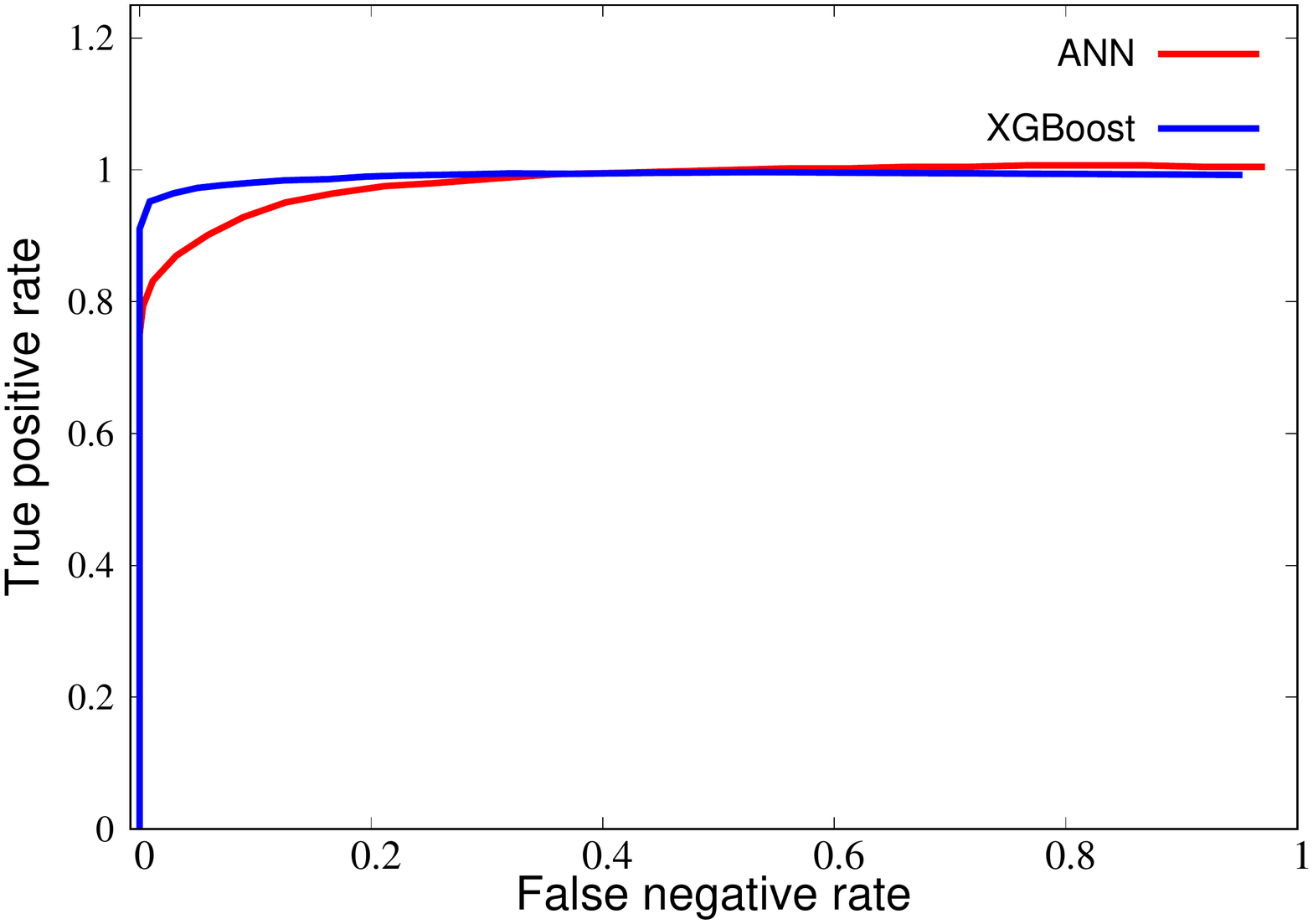}
\centering

\caption{ROC curves of BP 1 (top left), BP 2 (top right) and BP 3 (bottom centre) in case II with ANN and XGBoost.}
\label{rocww}
\end{figure}

In the gradient boosted decision tree analysis we have used 1000 estimators, maximum depth 2 and a learning rate 0.02.
 In case of ANN we have used four hidden layers with activation curve tanh and relu in succession, a batch-size 200 for each epoch, and 100 such epochs. For both XGBoost and ANN analysis we have used 80\% of the data for training and 20\% for test or validation of the algorithm. We found out that in case I, the invariant mass of the same-sign dilepton pair plays the most important role in signal-background identification, $m_{transverse}$, $\slashed{E_T}$, $p_T$ of the leading and sub-leading leptons being of relatively lower importance. In case II, the invariant mass of the lepton pair becomes less relevant as we have discussed earlier. The most important observable in this case turns out to be $m_{transverse}$ including the correlated ones, namely $m_{cluster}$, $m_{transverse}$ $H_T$ as seen in Figures~\ref{correlation}. 

In Figure~\ref{rocll} and~\ref{rocww} we present the Receiver Operating Characteristic (ROC) curves for all the benchmarks of case I and II. For different scenarios and benchmarks considered here, the area under the ROC curves vary within the range 0.92-0.99. This implies that it is indeed possible to gain high signal selection efficiency with extremely low background selection. One possible issue with this kind of analyses is the possibility of over-training, in which case the separation between signal and background becomes extremely good for the training sample but for the test sample it fails to achieve the same level of distinction. We have explicitly checked that in our case the algorithm is not over-training, as a result of which the area under the curves remain almost same for training and test sample. In Figure~\ref{rocll} we can see that the large signal selection efficiency ($\sim 90$\%) is achievable with extremely low background selection ($\sim 0.1$\%) in case of all the BPs. The invariant mass of dilepton pair is the major reason behind such separation. For clarity we have plotted the background selection rate (false positive rate) upto a smaller range in this figure. In Figure~\ref{rocww} we can see that for signal selection efficiency ($\sim 90$\%), one will have to allow $\sim 1$\% fake background in case of $W^{\pm}W^{\pm}$ final state. Evidently the results will worsen as compared to $\ell^{\pm}\ell^{\pm}$ final state. One can also see from Figures~\ref{rocll} and~\ref{rocww}, XGBoost performs slightly better than ANN in all cases, one deciding factor being the number of input variables~\cite{Roe:2004na}.

Next we compute the signal significance for all the benchmarks in case I and II with ANN and XGBoost. We present the results in Table~\ref{significance_ll_mlearning} and~\ref{significance_ww_mlearning} respectively. One can compare these results with the ones quoted in Table~\ref{significance_ll} and~\ref{significance_ww}. It is clear that in all cases there is significant improvement from rectangular cut-based analysis. We particularly point out the BP 2 and 3 in case II. In these two cases we observe striking improvement from the cut-based results. Finding the best possible combination of feature variables to separate the signal and background ANN helps us improve the significance. On the other hand XGBoost does the same by choosing the best possible set of cuts on the most relevant observables. We remark here that the data sample used by us for training purpose may in principle be subjected to some pre-assigned additional cuts, such as demanding specific invariant masses for same-sign dileptons. Such a practice usually improves the signal significance further\cite{Dey:2019lyr}. We have desisted from using such cuts, since the significance is already quite impressive.

\begin{table}[!hptb]
\begin{center}
\begin{footnotesize}
\begin{tabular}{| c | c | c | }
\hline
BP & $
{\cal S}$ (ANN) &  ${\cal S}$ (XGBoost)  \\
\hline
BP 1  &  5.9 $\sigma$ & 7.8 $\sigma$ \\
\hline
BP 2  & 9.3 $\sigma$ & 11.6 $\sigma$ \\
\hline
BP 3  & 6.4 $\sigma$ & 7.9 $\sigma$ \\
\hline
\end{tabular}
\end{footnotesize}
\caption{Signal significance for the benchmark points at 14 TeV with ${\cal L}$ = 3000 $fb^{-1}$ for case I with ANN and XGBoost. }
\label{significance_ll_mlearning}
\end{center}
\end{table}

\begin{table}[!hptb]
\begin{center}
\begin{footnotesize}
\begin{tabular}{| c | c | c | }
\hline
BP & $
{\cal S}$ (ANN) &  ${\cal S}$ (XGBoost)  \\
\hline
BP 1  &  3.6 $\sigma$ & 4.8 $\sigma$ \\
\hline
BP 2  & 3.9 $\sigma$ & 5.0 $\sigma$ \\
\hline
BP 3  & 3.4 $\sigma$ & 4.0 $\sigma$ \\
\hline
\end{tabular}
\end{footnotesize}
\caption{Signal significance for the benchmark points at 14 TeV with ${\cal L}$ = 3000 $fb^{-1}$ for case II with ANN and XGBoost. }
\label{significance_ww_mlearning}
\end{center}
\end{table}

\section{Conclusions}\label{sec7}

 We use the fact that theories with extended scalar sectors can provide viable candidates for DM portal, avoiding the constraints prevailing on the SM Higgs from direct search and relic density considerations. Keeping this in mind, we have explored the scenario where a CP-even scalar from a triplet acts as the portal to the dark sector, consistently with the role of the triplet in the Type-II seesaw mechanism for neutrino mass generation. One can find interesting regions of the parameter space, which are consistent with all the requirements from Higgs data, dark matter experiments, precision measurement as well as theoretical constraints. We have chosen a few representative benchmark points which give significant production cross-section for the heavy Higgs bosons as well as branching ratios in the invisible channel for the heavy CP-even scalar $H$. The production of $H$ along with doubly charged Higgs has the advantage of same-sign dilepton in the final state, which is a clean signal to look for at the LHC. We have considered two complimentary scenarios with low and high triplet VEV, and explored the reach of the high-luminosity LHC in probing both cases. We have found out that choosing suitable kinematical observables it is possible to achieve significant event rates in both channels for specific benchmark points. The region with low triplet VEV provides us better signal-background separation, having the advantage of invariant mass peak for the same-sign dileptons. The region with moderate to large triplet VEV do not have this invariant mass peak as a discriminating variable. Also this channel suffers from low leptonic branching of the $W$ bosons. We 
ameliorate such difficulties  by going beyond the rectangular cut-based analysis, applying 
gradient boosting as well as neural network techniques which strikingly improve the significance for all the scenarios. 

It has been already mentioned in Section~\ref{sec4} that the signals considered here can be mimicked by a situation where the
heavy triplet-dominated scalar $H$ has a substantial branching ratio into a pair of neutrinos,
something that can be envisioned for small values of the triplet VEV. In principle,
such a possibility can be distinguished by other collider signals of the Type-II
Seesaw scenario, and from a relatively detailed understanding of its parameter space
acquired thereby. In the (unlikely) case where such differentiation is impossible,
{\em searches for the signals suggested here will in any case serve to constrain  
a triplet DM portal}.

\section{Acknowledgement}

We thank Asesh Krishna Datta for valuable comments. This work was supported by funding available from the Department of Atomic Energy,  Government of India, for the Regional Centre for Accelerator-based Particle Physics (RECAPP), Harish-Chandra Research Institute. AD and JL thank Saha Institute of Nuclear Physics, Kolkata and Indian Institute of Science Education and Research Kolkata for hospitality, where substantial part of this work was done.  

\bibliographystyle{JHEP}
\bibliography{paperbib}

\end{document}